%% file: main.tex
\pdfoutput=1
\documentclass[11pt,a4paper]{article}

\usepackage[margin=1in]{geometry}
\usepackage{graphicx}
\graphicspath{{figures/}}

\usepackage{subcaption}
\usepackage{amsmath}
\usepackage{amssymb}
\usepackage{bm}
\usepackage{float}
\usepackage{algorithm}
\usepackage{algorithmic}
\usepackage{url}
\usepackage[colorlinks=true,allcolors=blue]{hyperref}
\usepackage[capitalise,noabbrev]{cleveref}
\usepackage{tikz}
\usetikzlibrary{arrows.meta,positioning}

\newcommand{\kyungtak}[1]{}

\newcommand{\gridx}{244}
\newcommand{\gridy}{324}
\newcommand{\gridz}{80}
\newcommand{\gridzsubset}{20}
\newcommand{\latentdim}{32}
\newcommand{\modesx}{122}
\newcommand{\modesy}{162}
\newcommand{\modesz}{10}
\newcommand{\repetitions}{2}

\newcommand{\predsteps}{4}

\newcommand{\trainparams}{40{,}540{,}324}
\newcommand{\trainparamsomega}{40{,}540{,}324}
\newcommand{\inferms}{123.64}
\newcommand{\channelratio}{0.25}

\newcommand{\papertitle}{Surrogate modeling of drift-reduced Braginskii turbulence with resistivity-conditioned Koopman neural operators}

\begin{document}

\title{\papertitle}
\author{%
  Ameir Shaa$^{1}$, Kyungtak Lim$^{1,*}$, Long Shan Chan$^{1}$, Claude Guet$^{1}$\\[0.5em]
  {\small $^{1}$School of Physical and Mathematical Sciences, Nanyang Technological University, Singapore 637371, Singapore}\\[0.25em]
  {\small $^{*}$Corresponding author}
}
\date{}

\maketitle

\begin{abstract}
Machine-learning-driven surrogate operators are developed for three-dimensional, nonlinear, flux-driven simulations of boundary plasma turbulence based on the two-fluid drift-reduced Braginskii model. Resistivity-conditioned Koopman neural operators (KNOs) are trained on Global Braginskii Solver (GBS) simulations, spanning low- to high-resistivity regimes. Separate fieldwise models are constructed for plasma density, electron temperature, electric potential, and vorticity. Evaluation at a held-out resistivity shows that the surrogates reproduce key short-horizon statistical features, including strong one-step agreement, spectral trends, and reduced pressure-gradient diagnostics. Field-dependent limitations remain, with vorticity showing the largest discrepancies and autoregressive rollout progressively departing from the reference simulation. The results demonstrate that resistivity-conditioned fieldwise neural operators provide useful fast emulators for selected boundary-plasma turbulence diagnostics, while stable long-horizon dynamical closure remains unresolved.
\end{abstract}

\noindent\textbf{Keywords:} Boundary plasma turbulence, magnetic confinement fusion, drift-reduced Braginskii model, Koopman neural operators, surrogate modeling

\input{sections/1_Intro}
\input{sections/2_Braginskii}
\input{sections/3_KNO}
\input{sections/4_Results}
\input{sections/5_Conclusion}

\section{Acknowledgments}
This work is partly funded by Ministry of Education (MOE) AcRF Tier 1 grants RS38/24 and RG178/25, and National Research Foundation Singapore (NRF) core funding ``Fusion Science for Clean Energy'', under the framework of Singapore Alliance with France for Fusion Energy (SAFE). The computational work for this article was performed on resources of the National Supercomputing Centre (NSCC), Singapore (Project 12003525).

\appendix
\input{sections/8_Appendix}

\bibliographystyle{ieeetr}
\bibliography{9_references}

\end{document}

%% file: sections/1_Intro.tex
\section{Introduction}
Anomalous cross-field transport remains a central limitation in magnetic-confinement fusion, degrading energy confinement through multiscale dynamics ranging from micro-instabilities to large intermittent transport events \cite{Connor2000, Garbet2010, Horton1999}. In the tokamak boundary, two-fluid models based on the drift-reduced Braginskii equations provide a tractable yet physically faithful description of these dynamics \cite{Braginskii1965, Zeiler1997}. Nevertheless, fully three-dimensional, nonlinear, flux-driven simulations remain computationally demanding, particularly when systematic scans over operating parameters such as plasma density, heating power, or magnetic geometry are required. This motivates machine-learning surrogate operators that can emulate selected observables across the parameter space without rerunning the full simulation.

Recent surrogate studies in plasma physics demonstrate the breadth of this effort, including generative models for Hasegawa–Wakatani turbulence \cite{Clavier2025}, interpolation surrogates for tokamak edge profiles \cite{Dasbach2023}, Fourier neural operator (FNO) surrogates for plasma evolution and edge simulations \cite{Gopakumar2024FNO,Gopakumar2025}, machine-learning-accelerated gyrokinetic workflows \cite{Narita2022}, and data-efficient learning methods from limited high-fidelity data and coarse-grid simulations \cite{Wan2025,Li2025}. These studies show that learned surrogates can reproduce useful statistical and parametric trends at a fraction of the simulation cost, also highlighting that surrogate fidelity depends strongly on the chosen diagnostic, the intended prediction horizon, and the explicit conditioning of physical control parameters when the dataset spans distinct turbulent regimes.

Leveraging the GBS surrogate framework in Ref.~\cite{Solheim2026}, which demonstrated proper-orthogonal-decomposition (POD)-based quasi-steady profile reconstruction in two-dimensional cross-sections, the present work extends data-driven modeling to the fully three-dimensional, time-dependent setting. We introduce resistivity-conditioned Koopman neural operators (KNOs) trained on electrostatic GBS simulations of drift-reduced Braginskii turbulence in toroidal geometry \cite{Krommes2002, Scott2005, Giacomin2022}. The scope is deliberately focused on whether a KNO surrogate can interpolate across a resistivity scan within a fixed reduced-fluid configuration, while preserving the short-horizon statistical signatures relevant to plasma-turbulence diagnostics. The surrogate is therefore treated as a fast conditional emulator for selected observables, rather than as a replacement for the underlying solver. Beyond diagnostic emulation, such low-cost three-dimensional predictions may also serve as approximate initial states for GBS restarts, bypassing the costly initial growth phase before quasi-steady turbulence~\cite{Solheim2026}.

Reduced-order modeling and neural operators provide a systematic route to learning state-to-state maps for PDE-governed dynamics \cite{Kaptanoglu2021,Vlasov2021,Li2021,Lu2021,Kovachki2023}. The Koopman framework is particularly attractive for nonlinear systems because it represents the dynamics in a lifted observable space where the evolution can be approximated as linear \cite{Koopman1931,Mezic2005,Mezic2013,Kutz2016,Williams2015,Tu2014,Arbabi2017}. Within this framework, KNOs offer a structured surrogate representation of nonlinear plasma evolution, allowing the dominant spatiotemporal dynamics to be learned efficiently in a latent space. In the present construction, the resistivity label $\nu_0$ is embedded directly into the Koopman mixing matrices, so that the learned evolution operator carries the parameter dependence. This provides a direct route to resistivity-conditioned interpolation across turbulent regimes.

\begin{figure*}[htbp]
\centering
\includegraphics[width=\textwidth]{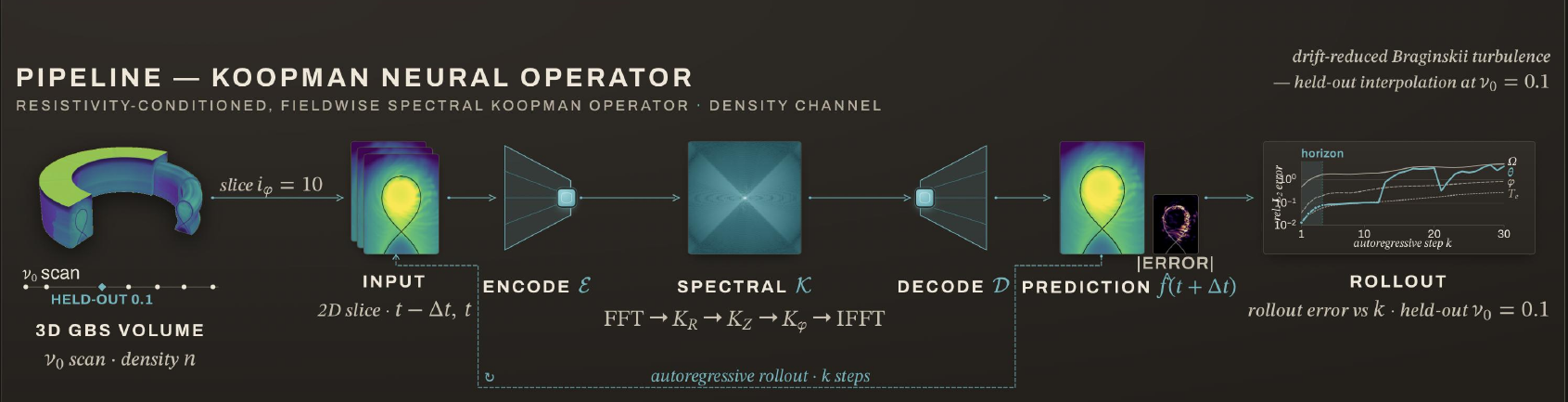}
\caption{Resistivity-conditioned Koopman neural operator workflow, illustrated for the density channel. A two-frame field history and the resistivity label $\nu_0$ are encoded, advanced in a latent Koopman space, and decoded to predict the next field snapshot. The same fieldwise workflow is used for all reported surrogate models.}
\label{fig:pipeline}
\end{figure*}

Separate KNO surrogates are trained for plasma density, electron temperature $T_e$, electric potential $\phi$, and vorticity $\Omega$ across a five-point resistivity scan ($\nu_0\in\{0,0.01,1,3,10\}$) and are evaluated at the held-out value $\nu_0=0.1$ as a controlled interpolation test. The results show strong one-step agreement for electron temperature and density, good preservation of the density spectral trend, and accurate recovery of reduced pressure-gradient structure, while vorticity remains the most challenging observable. At the same time, autoregressive rollout reveals that short-horizon statistical fidelity does not guarantee stable free-running emulation. Errors accumulate and spectral amplitudes drift as the model is iterated. The main contribution of this work is therefore to quantify where resistivity-conditioned fieldwise emulation succeeds at an unseen resistivity, and where long-horizon dynamical closure and coupled-field consistency remain unresolved. Figure~\ref{fig:pipeline} shows the resistivity-conditioned training and inference workflow, illustrated for the density channel and used consistently across the reported fieldwise surrogates.

The remainder of the paper is organized as follows. \cref{sec:braginskii_model} summarizes the drift-reduced Braginskii model, the GBS simulation setup, and the physical role of the resistivity label $\nu_0$. \cref{sec:kno_framework} introduces the resistivity-conditioned fieldwise KNO architecture, together with the training objective and data configuration. \cref{sec:validation} presents the held-out validation results, including pointwise, reduced-diagnostic, temporal, spectral, blob-detection, rollout, and computational-cost assessments. Finally, \cref{sec:conclusion} concludes the paper.

%% file: sections/2_Braginskii.tex
\section{Physical model and simulation data}\label{sec:braginskii_model}
The reference data are generated with the three-dimensional drift-reduced Braginskii equations \cite{Zeiler1997}, as implemented in the GBS code \cite{Ricci2012,Giacomin2022}. GBS evolves a nonlinear two-fluid model for boundary plasma turbulence in a non-field-aligned cylindrical coordinate system, making it suitable for surrogate studies under controlled variations of physical parameters. In the present work, we consider electrostatic simulations in an axisymmetric single-null magnetic equilibrium and use them as reference trajectories for training and validating resistivity-conditioned surrogate models. The plasma variables considered here are collected as
\begin{align}
 \mathbf{q}(t,\mathbf{r}) = \bigl(n,\;\Omega, \;v_{\parallel i},\;v_{\parallel e},\;T_i,\;T_e, \; \phi \bigr),
 \label{eq:state_vector}
\end{align}
where $n$ is the plasma density, $\Omega$ is the scalar vorticity, $v_{\parallel i}$ and $v_{\parallel e}$ are the ion and electron parallel velocities, and $T_i$ and $T_e$ are the ion and electron temperatures. The ion and electron pressures are $p_i=n\tau T_i$ and $p_e=nT_e$, with $\tau=T_{i0}/T_{e0}$.

Under these assumptions, the GBS model is written as
\begin{align}
 \frac{\partial n}{\partial t} &= -\frac{\rho_*^{-1}}{B}[\phi, n] + \frac{2}{B}\bigl[C(p_e)-nC(\phi)\bigr]-\nabla_\parallel (nv_{\parallel e}) + D_n \nabla^2_\perp n + s_n,
\label{GBS_density}\\[2pt]
 \frac{\partial \Omega}{\partial t} &= -\frac{\rho_*^{-1}}{B} \nabla \cdot [\phi, \boldsymbol{\omega}] -\nabla \cdot \bigl(v_{\parallel i} \nabla_\parallel\boldsymbol{\omega}\bigr) + B^2 \nabla_\parallel j_\parallel + 2B C(p_e + \tau p_i) + \frac{B}{3}C(G_i) + D_\Omega \nabla_\perp^2 \Omega,
\label{GBS_vorticity}\\[2pt]
 \frac{\partial v_{\parallel i}}{\partial t} &= -\frac{\rho_*^{-1}}{B}[\phi, v_{\parallel i}]-v_{\parallel i} \nabla_\parallel v_{\parallel i} - \frac{1}{n}\nabla_\parallel (p_e + \tau p_i) -\frac{2}{3n}\nabla_\parallel G_i + D_{v_{\parallel i}}\nabla_\perp^2 v_{\parallel i},
\label{GBS_vpi}\\[2pt]
 \frac{\partial v_{\parallel e}}{\partial t} &= -\frac{\rho_*^{-1}}{B}[\phi, v_{\parallel e}] - v_{\parallel e}\nabla_\parallel v_{\parallel e} + \frac{m_i}{m_e}\Bigl(\nu j_\parallel + \nabla_\parallel \phi -\tfrac{1}{n}\nabla_\parallel p_e-0.71 \nabla_\parallel T_e -\tfrac{2}{3n}\nabla_\parallel G_e \Bigr) \nonumber \\
 &\quad + D_{v_{\parallel e}}\nabla_\perp^2 v_{\parallel e}, 
 \label{GBS_ve}\\[2pt]
 \frac{\partial T_i}{\partial t} &= -\frac{\rho_*^{-1}}{B}[\phi, T_i] - v_{\parallel i}\nabla_\parallel T_i + \frac{4}{3}\frac{T_i}{B}\Bigl[C(T_e) + \tfrac{T_e}{n}C(n)-C(\phi) \Bigr] -\frac{10}{3}\tau \frac{T_i}{B}C(T_i) \nonumber \\
 &\quad + \frac{2}{3}T_i\Bigl[(v_{\parallel i}-v_{\parallel e})\tfrac{\nabla_\parallel n}{n}-\nabla_\parallel v_{\parallel e}\Bigr] + 2.61\,\nu n (T_e -\tau T_i)+ \nabla_\parallel( \chi_{\parallel i}\nabla_\parallel T_i) + D_{T_i}\nabla_\perp^2 T_i + s_{T_i},
\label{GBS_Ti}\\[2pt]
 \frac{\partial T_e}{\partial t} &= -\frac{\rho_*^{-1}}{B}[\phi, T_e] - v_{\parallel e}\nabla_\parallel T_e + \frac{2}{3}T_e\Bigl[0.71\, \tfrac{\nabla_\parallel j_\parallel}{n}-\nabla_\parallel v_{\parallel e}\Bigr] -2.61\,\nu n (T_e-\tau T_i) \nonumber \\
 &\quad + \frac{4}{3}\frac{T_e}{B}\Bigl[\tfrac{7}{2}C(T_e)+\tfrac{T_e}{n}C(n)-C(\phi) \Bigr] + \nabla_\parallel (\chi_{\parallel e}\nabla_\parallel T_e) + D_{T_e}\nabla_\perp^2 T_e + s_{T_e}. \label{GBS_electron_temperature}
\end{align}
The electrostatic potential is obtained from the generalized Poisson equation
\begin{align}
 \nabla \cdot \bigl( n\nabla_\perp \phi \bigr) = \Omega - \tau \nabla_\perp^2 p_i.
\label{GBS_Poisson}
\end{align}
Here $\boldsymbol{\omega}=n\nabla_\perp\phi+\tau\nabla_\perp p_i$ is the generalized vorticity vector, $\Omega=\nabla\cdot\boldsymbol{\omega}$ is the scalar vorticity, $j_\parallel=n(v_{\parallel i}-v_{\parallel e})$ is the parallel current, $[f,g]=\mathbf{b}\cdot(\nabla f\times\nabla g)$ denotes the Poisson bracket associated with $E\times B$ advection, and $C(\cdot)$ is the magnetic-curvature operator. The system includes the main physical ingredients required for boundary-plasma turbulence, such as $E\times B$ advection, curvature drive, parallel dynamics, resistive electron response, thermal conduction, sources, and perpendicular diffusion. Since this work focuses on surrogate-model construction and validation rather than on rederiving the GBS formulation, details of the numerical scheme, source profiles, and boundary conditions are not repeated here and can be found in Refs.~\cite{Ricci2012,Giacomin2022}.

Density and temperature sources are applied inside the last closed flux surface, representing the combined effect of ionisation and ohmic heating that sustains the flux-driven turbulence: the density source is a Gaussian in the equilibrium flux function $\psi(R,Z)$ and the temperature source a $\tanh$ profile in $\psi$, with fixed amplitudes and radial widths. The magnetic equilibrium is a single-null configuration generated by an external poloidal-field coil set, held fixed across the resistivity scan, so all simulations share the same geometry and source placement relative to the last closed flux surface~\cite{Giacomin2022}.

All quantities follow the standard GBS normalization. Density is normalized to a reference density $n_0$, ion and electron temperatures to $T_{i0}$ and $T_{e0}$, parallel velocities to the reference sound speed $c_{s0}=\sqrt{T_{e0}/m_i}$, and the electrostatic potential to $T_{e0}/e$. Perpendicular lengths are normalized to the ion sound Larmor radius $\rho_{s0}=c_{s0}/\Omega_{ci}$, parallel lengths to the tokamak major radius $R_0$, and time to $R_0/c_{s0}$. 

\begin{figure}[H]
\centering
\includegraphics[width=\textwidth]{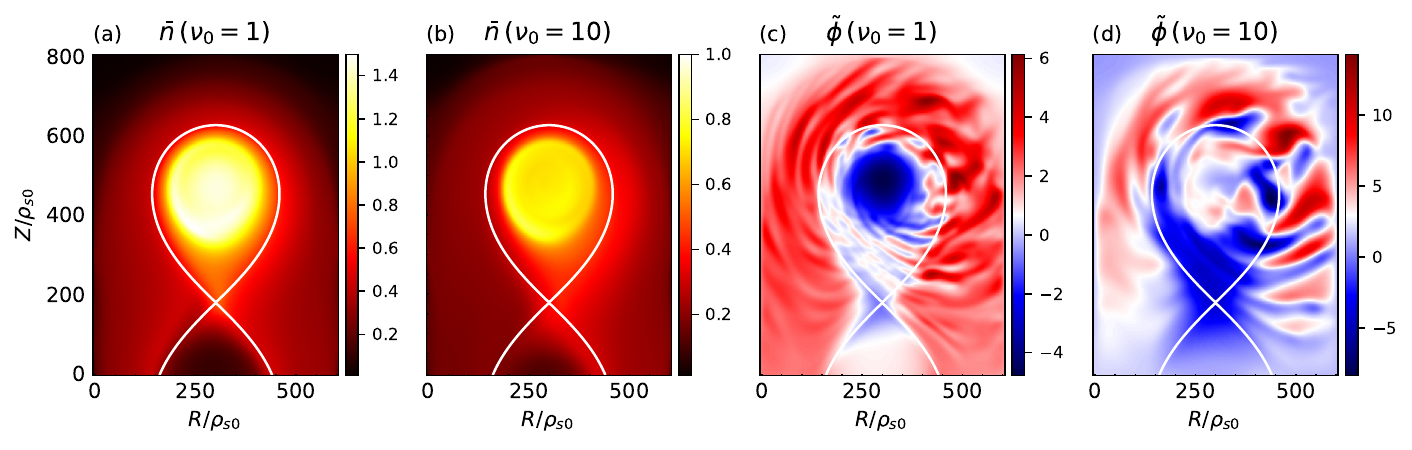}
\caption{
Two-dimensional snapshots from GBS simulations at different resistivities. Panels (a,b) show the mean density $\bar{n}$, averaged over the toroidal angle and a $10\,t_0$ (where $t_0 = R_0/c_{s0}$) time window in the quasi-steady phase (after initial transients have decayed), for $\nu_0=1$ and $\nu_0=10$, while panels (c,d) show the corresponding electric potential fluctuations $\tilde{\phi}$. The white contour denotes the separatrix, and $R$ and $Z$ are normalized to $\rho_{s0}$.}
\label{fig:GBS2D}
\end{figure}

A key control parameter in this study is the normalized Spitzer resistivity
\begin{align}
\nu = \nu_0 T_e^{-3/2},
\label{eq:resistivity}
\end{align}
where $\nu_0$ denotes the nominal resistivity label assigned to each simulation. In the normalized GBS formulation, $\nu_0$ is proportional to the reference density $n_0$ when the other reference quantities are fixed. Therefore, scanning $\nu_0$ from low to high values corresponds to moving from lower- to higher-density boundary-plasma regimes. As shown in Figure~\ref{fig:GBS2D}, the resistivity scan modifies both the mean profiles and fluctuation activity, showing a more peaked density profile and weaker electrostatic-potential fluctuations in the lower-resistivity case, whereas the higher-resistivity case exhibits a broader density profile and stronger fluctuation activity. This makes $\nu_0$ a natural conditioning variable for testing surrogate interpolation across resistivity-dependent turbulent states.

The surrogate is trained on GBS simulations with
\begin{align}
\;\;\nu_0 \in \{0,\; 0.01,\; 1,\; 3,\; 10\},
\label{eq:training_resistivities}
\end{align}
and evaluated at the held-out value $\nu_0=0.1$, with $\nu_0=0.5$ used for validation-based checkpoint selection. The test case therefore probes interpolation in the resistivity label within the same reduced-fluid configuration, rather than extrapolation beyond the training range. All other input parameters are kept fixed in the present study, so that the role of resistivity in the learned dynamics can be isolated.

GBS stores the simulation output on a three-dimensional grid with $(n_R\times n_Z\times n_\varphi)=(\gridx\times\gridy\times\gridz)$. The full simulation state can be written as
\begin{align}
\mathcal{X}(t,\mathbf{r}) \in \mathbb{R}^{8\times n_R\times n_Z\times n_\varphi},
\label{eq:full_state_tensor}
\end{align}
where the eight stored channels are the seven evolved fields---\texttt{theta} ($\theta=\ln n$, log-density), \texttt{omega} ($\Omega$, vorticity), \texttt{vpari} ($v_{\parallel i}$, ion parallel velocity), \texttt{vpare} ($v_{\parallel e}$, electron parallel velocity), \texttt{temperaturi} ($T_i$, ion temperature), \texttt{temperature} ($T_e$, electron temperature), and \texttt{strmf} ($\phi$, electrostatic potential)---together with the magnetic flux \texttt{psi} ($\psi$) as an additional diagnostic quantity. In this work, the surrogate is not trained to learn a coupled map for the full GBS state vector. Instead, separate fieldwise models are constructed for selected scalar observables: plasma density, electron temperature, electric potential, and vorticity. The full GBS simulations then provide the reference dynamics against which the resistivity-conditioned surrogate predictions are evaluated.

%% file: sections/3_KNO.tex
\section{Resistivity-conditioned Koopman neural operator}\label{sec:kno_framework}
The surrogate model is formulated for selected diagnostic fields rather than the full Braginskii state vector. We train independent fieldwise models for the learned density variable $\theta=\ln n$, electron temperature $T_e$, electric potential $\phi$, and vorticity $\Omega$. This fieldwise formulation reduces the dimensionality of the learning problem and allows the fidelity of each physical observable to be assessed independently. 

For a target field $f(\mathbf{x},t;\nu_0)$, the model learns a short-time evolution map,
\begin{align}
\mathcal{G}_f:\bigl(f(\mathbf{x},t-\Delta t),\,f(\mathbf{x},t),\,\nu_0\bigr)\mapsto \hat{f}(\mathbf{x},t+\Delta t).
\label{eq:kno_fieldwise_operator}
\end{align}
Here $f(\mathbf{x},t)$ denotes a three-dimensional field snapshot on the GBS grid, and $\nu_0$ labels the resistivity regime of the corresponding simulation. The model, therefore, acts as a resistivity-conditioned emulator of the local-in-time field evolution. During training, a single forward pass jointly predicts the next four frames and the loss is averaged over all four horizons, providing multi-horizon supervision over a short window rather than autoregressive feedback, while the validation metrics reported in \cref{sec:validation} use only the first predicted frame.

The architecture follows the Koopman operator framework, in which nonlinear dynamics are represented by approximately linear evolution in a lifted observable space \cite{Koopman1931,Mezic2005,Mezic2013,Kutz2016,Williams2015,Tu2014,Arbabi2017}. Building on Koopman-based studies of nonlinear plasma dynamics \cite{Faraji2025}, we extend this idea to a resistivity-conditioned, fieldwise neural operator for three-dimensional boundary plasma. Unlike standard FNO surrogates \cite{Gopakumar2024FNO,Gopakumar2025}, where physical parameters enter through the input channels or through the training distribution, the present model embeds the resistivity label $\nu_0$ directly into the latent Koopman evolution operator. In this way, the learned evolution operator itself carries the resistivity dependence.

\subsection{Fieldwise latent evolution}

The input history is first encoded into a latent representation. At each grid point, the two most recent values of the target field are collected as
\begin{align}
\mathbf{h}_f(\mathbf{x},t)=\bigl(f(\mathbf{x},t-\Delta t),\,f(\mathbf{x},t)\bigr),
\label{eq:temporal_history_vector}
\end{align}
and mapped to a local latent vector by a temporal encoder,
\begin{align}
\mathbf{z}(\mathbf{x},t)=\tanh\left[\mathcal{E}_{\mathrm{temp}}\bigl(\mathbf{h}_f(\mathbf{x},t)\bigr)\right].
\label{eq:temporal_encoder}
\end{align}
A three-dimensional spatial encoder then maps $\mathbf{z}$ to an initial latent field,
\begin{align}
\mathbf{Z}^{(0)}=\mathcal{E}_{\mathrm{3D}}(\mathbf{z}).
\label{eq:3D}
\end{align}
This latent field contains both the short-time history and the spatial structure needed to advance the target field by one step.

The resistivity dependence is introduced in the latent evolution operator. In Fourier space, the direction-wise Koopman operator is written as the sum of a baseline contribution, a resistivity-dependent correction, and a timestep-dependent correction,
\begin{align}
\mathcal{K}_d(\nu_0,\Delta t)=\mathcal{K}_d^{(0)}+\mathcal{K}_d^{(\nu)}(\nu_0)+\mathcal{K}_d^{(\Delta t)}(\Delta t), \qquad d\in\{R,Z,\varphi\}.
\label{eq:Kd_additive}
\end{align}
Here $\mathcal{K}_d^{(0)}$ is a baseline operator, while $\mathcal{K}_d^{(\nu)}$ and $\mathcal{K}_d^{(\Delta t)}$ are learned corrections associated with the resistivity label and timestep. Thus, $\nu_0$ does not simply enter as an additional input channel. Instead, it modifies the latent evolution operator, so that different resistivity regimes correspond to different learned short-time dynamics within the same architecture. This is the main distinction from standard FNO-type surrogates, where parameter dependence is typically supplied through the input data or the training distribution \cite{Gopakumar2024FNO,Gopakumar2025}.

The spectral update is applied sequentially along the retained $R$, $Z$, and $\varphi$ Fourier directions. In compact form, the latent field is advanced by
\begin{align}
 \mathbf{Z}^{(n+1)} = \mathbf{Z}^{(n)} + \mathcal{K}\bigl(\mathbf{Z}^{(n)},\nu_0,\Delta t\bigr), \qquad n = 0,\ldots,N_{\mathrm{iter}}-1,
 \label{eq:kno_residual_update}
\end{align}
where $\mathcal{K}$ denotes the composed Fourier-space update followed by the inverse transform back to physical space. The residual form is used because the surrogate predicts a short-time correction to the latent state rather than replacing the full latent representation in one step. In the reported models, $N_{\mathrm{iter}}=2$ residual Koopman updates are used, and $\modesx\times\modesy\times\modesz$ Fourier modes are retained. The $\modesx\times\modesy\times\modesz$ mode cube ($\modesx=\gridx/2$, $\modesy=\gridy/2$, $\modesz=\gridzsubset/2$) follows the \texttt{torch.fft.rfftn} convention on $(R,Z,\varphi)$ and does not impose an extra spectral cut along $R$ and $Z$. Training uses a $\gridzsubset$-plane contiguous toroidal subset of the full $\gridz$-plane GBS grid, with the full torus recovered spectrally as in Appendix~\ref{sec:z-recon}.

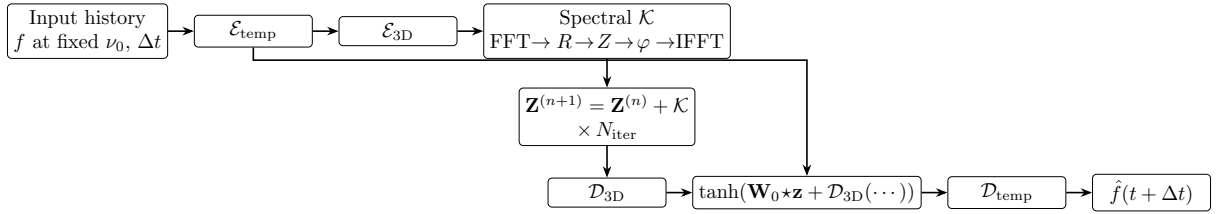
\begin{figure}[htbp]
\centering
\resizebox{\linewidth}{!}{%
\begin{tikzpicture}[
 font=\small,
 box/.style={draw, rounded corners=2pt, align=center, inner sep=3pt,
 minimum width=2.05cm},
 arr/.style={-{Stealth[length=2mm]}, thick}
]
\node[box] (in) {Input history\\$f$ at fixed $\nu_0$, $\Delta t$};
\node[box, right=0.45cm of in] (et) {$\mathcal{E}_{\mathrm{temp}}$};
\node[box, right=0.45cm of et] (e3) {$\mathcal{E}_{\mathrm{3D}}$};
\node[box, right=0.45cm of e3] (spec) {Spectral $\mathcal{K}$\\
 FFT$\to R\!\to\!Z\!\to\!\varphi\to$IFFT};
\node[box, below=0.5cm of spec] (res) {$\mathbf{Z}^{(n+1)}=\mathbf{Z}^{(n)}
 +\mathcal{K}$\\$\times\,N_{\mathrm{iter}}$};
\node[box, below=0.55cm of res] (d3) {$\mathcal{D}_{\mathrm{3D}}$};
\node[box, right=0.45cm of d3] (skip) {$\tanh(\mathbf{W}_0\!\star\!\mathbf{z}
 +\mathcal{D}_{\mathrm{3D}}(\cdots))$};
\node[box, right=0.45cm of skip] (dec) {$\mathcal{D}_{\mathrm{temp}}$};
\node[box, right=0.45cm of dec] (out) {$\hat{f}(t+\Delta t)$};
\draw[arr] (in) -- (et);
\draw[arr] (et) -- (e3);
\draw[arr] (e3) -- (spec);
\draw[arr] (spec) -- (res);
\draw[arr] (res) -- (d3);
\draw[arr] (d3) -- (skip);
\draw[arr] (et.south) |- ++(0.15,-0.22) -| (skip.north);
\draw[arr] (skip) -- (dec);
\draw[arr] (dec) -- (out);
\end{tikzpicture}%
}
\caption{Fieldwise resistivity-conditioned KNO workflow. A two-frame history of a target field is encoded, lifted to a three-dimensional latent field, evolved by a Koopman operator conditioned on $\nu_0$, and decoded to the next field snapshot. The resistivity label enters through the latent evolution operator rather than through the encoder or decoder.}
\label{fig:kno_arch}
\end{figure}

After the latent evolution, the spatially decoded Koopman correction is combined with the temporally encoded latent and passed through a pointwise $\tanh$ nonlinearity,
\begin{align}
 \mathbf{z}^{\mathrm{final}} &= \tanh\bigl(\mathbf{W}_0 \star \mathbf{z} + \mathcal{D}_{\mathrm{3D}}(\mathbf{Z}^{(N_{\mathrm{iter}})})\bigr), \label{eq:skip}\\
 \hat{f}(t+\Delta t) &= \mathcal{D}_{\mathrm{temp}}(\mathbf{z}^{\mathrm{final}}), \label{eq:kno_decode}
\end{align}
where $\mathbf{W}_0$ is a learned $1\times1\times1$ convolution that mixes latent channels and $\star$ denotes that convolution. The skip retains information from the original two-frame history, while the Koopman block supplies the nonlocal spatial correction. The overall workflow is summarized in \cref{fig:kno_arch}.

For reproducibility, the fieldwise forward pass is summarized in Algorithm~\ref{alg:kno_forward}. The algorithm provides a schematic description of the model operation, in which a two-frame field history is encoded, evolved in a resistivity-conditioned latent Koopman space, and decoded into a short forecast window. The first predicted frame is used for the one-step validation metrics, while all predicted frames enter the training loss.

\begin{algorithm}[htbp]
\caption{Fieldwise KNO forward pass}
\label{alg:kno_forward}
\begin{algorithmic}[1]
\REQUIRE Field history $\{f(\mathbf{x},t_j)\}_{j=0}^{H-1}$, resistivity label $\nu_0$, timestep $\Delta t$, number of Koopman updates $N_{\mathrm{iter}}$
\ENSURE Predicted sequence $\{\hat{f}(\mathbf{x},t+s\Delta t)\}_{s=1}^{N_{\mathrm{pred}}}$

\STATE $\mathbf{z} \leftarrow \tanh\bigl[\mathcal{E}_{\mathrm{temp}}\bigl(\{f(\mathbf{x},t_j)\}_{j=0}^{H-1}\bigr)\bigr]$ \COMMENT{encode local temporal history}

\STATE $\mathbf{Z}^{(0)} \leftarrow \mathcal{E}_{\mathrm{3D}}(\mathbf{z})$ \COMMENT{lift to spatial latent field}

\FOR{$n=0$ \TO $N_{\mathrm{iter}}-1$}
 \STATE $\tilde{\mathbf{Z}} \leftarrow \mathrm{FFT}\bigl[\mathbf{Z}^{(n)}\bigr]$
 \FOR{$d \in \{R, Z, \varphi\}$}
 \STATE $\mathcal{K}_d \leftarrow \mathcal{K}_d^{(0)}+\mathcal{K}_d^{(\nu)}(\nu_0)+\mathcal{K}_d^{(\Delta t)}(\Delta t)$
 \STATE $\tilde{\mathbf{Z}} \leftarrow \mathcal{K}_d\,\tilde{\mathbf{Z}}$ \COMMENT{direction-wise spectral mixing}
 \ENDFOR
 \STATE $\mathbf{Z}^{(n+1)} \leftarrow \mathbf{Z}^{(n)}+\mathrm{IFFT}\bigl[\tilde{\mathbf{Z}}\bigr]$ \COMMENT{residual latent update}
\ENDFOR

\STATE $\mathbf{z}^{\mathrm{final}} \leftarrow \tanh\bigl(\mathbf{W}_0 \star \mathbf{z}+\mathcal{D}_{\mathrm{3D}}\bigl(\mathbf{Z}^{(N_{\mathrm{iter}})}\bigr)\bigr)$ \COMMENT{spatial decode, then skip}

\STATE $\{\hat{f}(\mathbf{x},t+s\Delta t)\}_{s=1}^{N_{\mathrm{pred}}} \leftarrow \mathcal{D}_{\mathrm{temp}}\bigl(\mathbf{z}^{\mathrm{final}}\bigr)$

\RETURN $\{\hat{f}(\mathbf{x},t+s\Delta t)\}_{s=1}^{N_{\mathrm{pred}}}$
\end{algorithmic}
\end{algorithm}

In the reported configuration, $H=2$ (the number of input time frames), $N_{\mathrm{iter}}=2$, and $N_{\mathrm{pred}}=4$. Training samples are drawn from the late, quasi-steady turbulent phase of the trajectories. The one-step diagnostics in \cref{sec:validation} use $\hat{f}(\mathbf{x},t+\Delta t)$, while the full four-frame prediction window is used in the training loss.

\subsection{Training objective and data configuration}
The model is trained by minimizing a weighted combination of a decoded prediction loss and auxiliary latent-consistency terms,
\begin{align}
 \mathcal{L}_{\mathrm{total}} = 5\,\mathcal{L}_{\mathrm{pred}} + 0.5\,\mathcal{L}_{\mathrm{aux}}.
 \label{eq:Ltotal}
\end{align}

The primary prediction loss is
\begin{equation}
 \mathcal{L}_{\mathrm{pred}} = \mathrm{MSE}\bigl(\hat{\mathbf{f}}_{1:\predsteps},\,\mathbf{f}^{\mathrm{target}}_{1:\predsteps}\bigr),
 \label{eq:Lpred}
\end{equation}
where $\hat{\mathbf{f}}_{1:\predsteps}=\{\hat{f}(\mathbf{x},t+s\Delta t)\}_{s=1}^{\predsteps}$ is the predicted multi-frame sequence, $\mathbf{f}^{\mathrm{target}}_{1:\predsteps}$ is the corresponding GBS target sequence, and $\mathrm{MSE}$ denotes the mean over all retained grid points and prediction steps. The auxiliary loss $\mathcal{L}_{\mathrm{aux}}$ contains one additional prediction route and three encode--decode consistency terms applied before and after the Koopman stack. These auxiliary terms improve training stability by encouraging consistent encode–decode behavior in latent space; they are not constraints derived from the Braginskii equations. The auxiliary loss terms are detailed in full in \cref{sec:appendix-impl}.

Each target field is standardized using the training-set mean and standard deviation
\begin{align}
 f_{\mathrm{std}} = \frac{f-\mu_f}{\sigma_f},
 \label{eq:fieldwise-normalization}
\end{align}
while $\nu_0$ and the time coordinate are passed in their native normalized GBS (simulation) units rather than being re-standardized; in particular $\nu_0$ is the normalized (Spitzer) resistivity label. For each trajectory, the available time samples are divided into training, validation, and testing subsets using a $70/15/15$ temporal split. The full optimization settings are listed in \cref{tab:optimization}.

The fieldwise models are trained on GBS simulations with $\nu_0 \in \{0,0.01,1,3,10\}$ and evaluated at the held-out value $\nu_0=0.1$. Thus, the main validation case tests interpolation within the resistivity range rather than extrapolation outside the training domain. The case $\nu_0=0.5$ is used only for validation-based checkpoint selection and is not included in the main validation figures. The reference data are defined on a $\gridx\times\gridy\times\gridz$ structured grid in $(R,Z,\varphi)$. The surrogate is trained and evaluated on a retained contiguous 20-plane toroidal subset of the same poloidal mesh. Unless otherwise stated, the representative toroidal plane $i_\varphi=10$ is used for the poloidal-slice diagnostics in \cref{sec:validation}. Code field identifiers are listed in \cref{sec:appendix-impl}.

\begin{table}[htbp]
\centering
\caption{Optimization settings for all reported fieldwise KNO runs.}
\label{tab:optimization}
\begin{tabular}{ll}
\hline
Setting & Value \\
\hline
Optimizer & Adam \cite{Kingma2015} \\
Learning rate & $10^{-3}$ \\
LR schedule & Step decay, factor 0.8 every 500 epochs \\
Weight decay & $10^{-4}$ \\
Batch size & 8 for $\Omega$, $\phi$, $T_e$ and 4 for $\theta$ \\
Maximum epochs & 20000 \\
Loss weights & $\lambda_{\mathrm{pred}}=5$ and $\lambda_{\mathrm{aux}}=0.5$ \\
\hline
\end{tabular}
\end{table}

\begin{figure}[H]
\centering
\includegraphics[width=\linewidth]{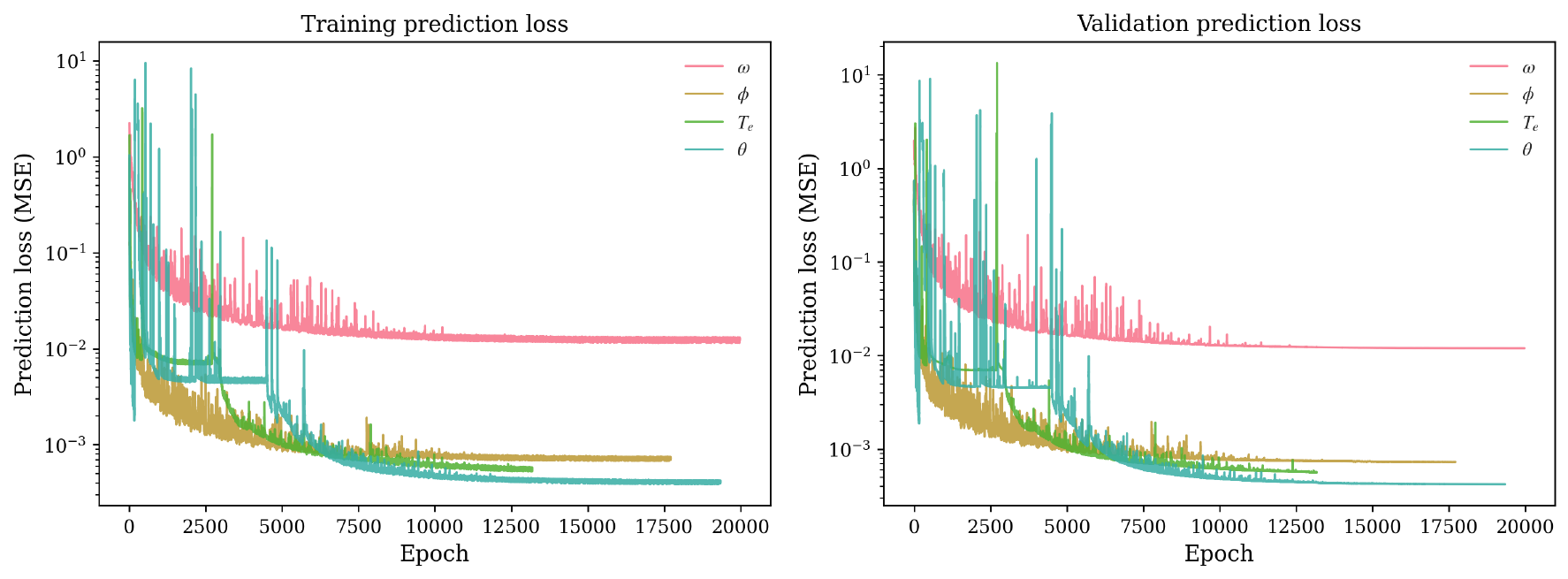}
\caption{Training and within-trajectory validation prediction loss versus epoch for the four fieldwise surrogates. The validation curve is obtained from the temporal validation split of each training-$\nu_0$ trajectory. Loss spikes coincide with the step-decay learning-rate schedule applied every 500 epochs and are most pronounced for vorticity, consistent with its higher spatial complexity. The validation-based checkpoint-selection case at $\nu_0=0.5$ is not shown.}
\label{fig:loss_curves}
\end{figure}

%% file: sections/4_Results.tex
\section{Validation and Statistical Fidelity}\label{sec:validation}
This section evaluates the resistivity-conditioned fieldwise KNO surrogates against the GBS reference at the held-out resistivity $\nu_0=0.1$. The validation is organized from local short-horizon accuracy to increasingly physics-oriented and dynamical diagnostics. We first assess one-step field reconstruction on a representative poloidal slice, followed by the three-dimensional density prediction, reduced pressure-gradient structure, teacher-forced temporal statistics, spectral content, blob detection, autoregressive rollout stability, and inference cost. This ordering separates conditional short-horizon statistical fidelity, where the surrogate remains close to the reference trajectory, from the more stringent requirement of autonomous long-horizon stability.

\subsection{Pointwise fidelity}
\begin{figure}[H]
\centering
\includegraphics[width=\linewidth]{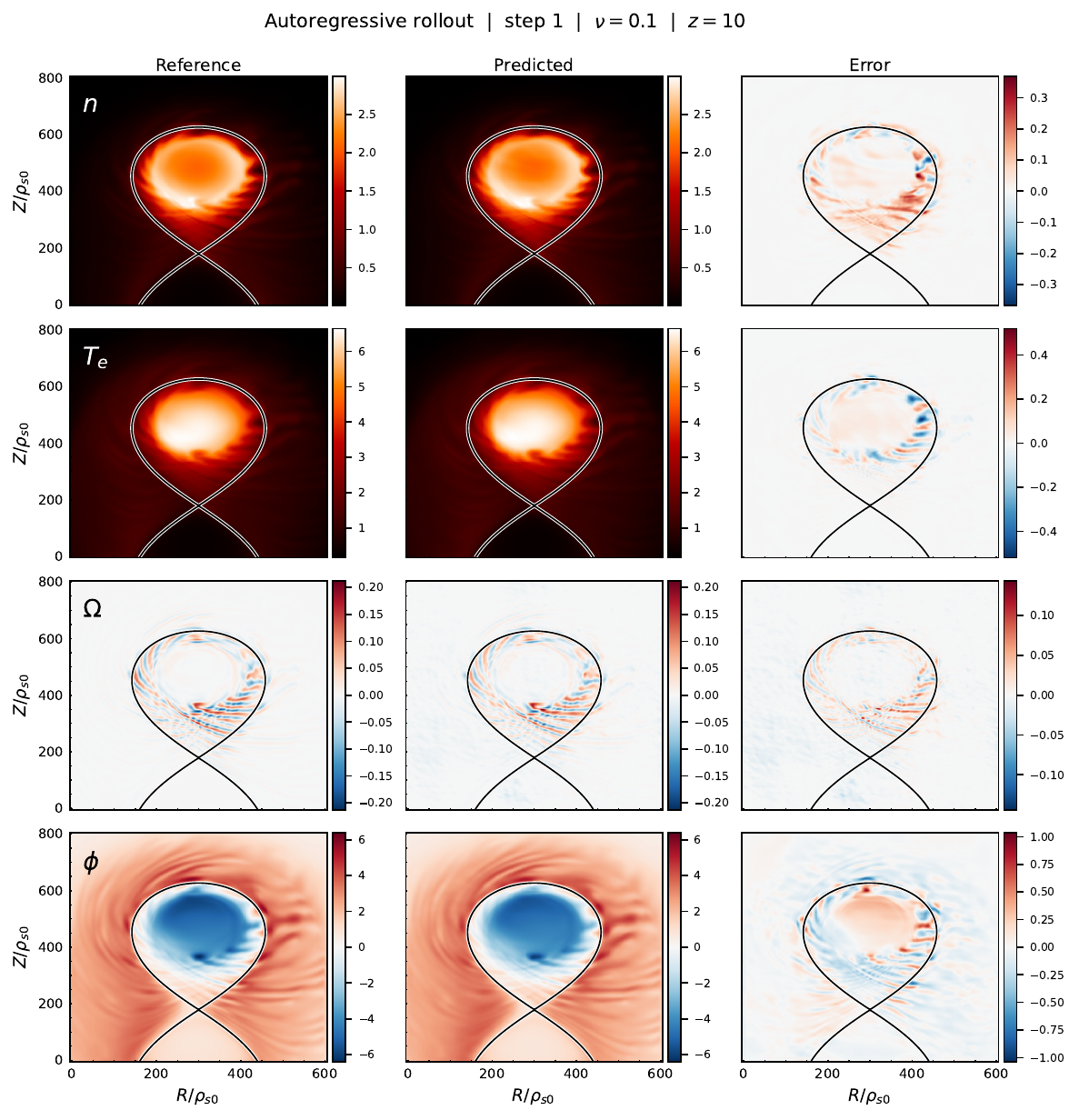}
\caption{Held-out interpolation at $\nu_0=0.1$ on the poloidal slice at $i_\varphi=10$. Rows show, from top to bottom, plasma density $n$, electron temperature $T_e$, vorticity $\Omega$, and electric potential $\phi$. Columns show the GBS reference, the KNO prediction, and the absolute error. The separatrix is represented as solid contours.}
\label{fig:2Dprediction}
\end{figure}

We first assess whether the fieldwise KNOs reproduce the GBS state at the omitted resistivity value $\nu_0=0.1$. Table~\ref{tab:fieldwise-summary} summarizes one-step slice metrics at the representative toroidal plane $i_\varphi=10$, while Figure~\ref{fig:2Dprediction} compares the GBS reference, KNO prediction, and absolute error for each field. Here the forecast interval is the spacing between consecutive saved GBS frames, $\Delta t_{\mathrm{out}}=0.1$ (in $R_0/c_{s0}$), so the one-step prediction maps $t\to t+\Delta t_{\mathrm{out}}$. As a reference, the persistence prediction $\hat x(t+\Delta t_{\mathrm{out}})=x(t)$, whose error equals the inter-frame change, is reported in the last column of Table~\ref{tab:fieldwise-summary}: the KNO one-step error is below the inter-frame change for every field.

Figure~\ref{fig:2Dprediction} shows that the density and electron temperature fields are reproduced with high pointwise fidelity across both the core and SOL regions, consistent with their near-unity $R^2$ values in Table~\ref{tab:fieldwise-summary}. The largest residuals are localized near steep-gradient regions, particularly around the separatrix and inner boundary. The electric potential is also well reconstructed, although its errors show coherent structure along the separatrix, indicating sensitivity to the phase of the large-scale $\mathbf{E}{\times}\mathbf{B}$ pattern rather than a global amplitude mismatch. Vorticity remains the most challenging field, with the largest absolute errors concentrated in fine-scale filamentary structures, where derivative sensitivity is strongest.

\input{results/fieldwise_summary_table.tex}

The one-step metrics confirm this field-dependent hierarchy. Electron temperature achieves near-unity agreement, with $R^2>0.999$, while the density and electric potential remain highly accurate with $R^2>0.99$. Vorticity shows the lowest pointwise fidelity, with $R^2\simeq0.76$ and the largest relative $L_2$ error. This ranking is physically expected because $T_e$ and $n$ are relatively smooth, whereas vorticity emphasizes small-scale derivative structure and is therefore more sensitive to high-wavenumber errors. 

The vorticity error should also be interpreted in view of the fieldwise training strategy. In the electrostatic Braginskii system, $\Omega$ and $\phi$ are coupled through the generalized polarization relation~\eqref{GBS_Poisson}, in which $\Omega$ follows from density-weighted perpendicular derivatives of $\phi$. This derivative coupling enhances sensitivity to high-$k_\perp$ content, so modest spatial errors in $\phi$ are amplified in $\Omega$. Because the present surrogates predict $\Omega$ and $\phi$ independently, this polarization consistency is not enforced during inference. This single limitation underlies both the instantaneous vorticity error here and the spectral and stability results discussed below.

\begin{figure}[H]
\centering
\includegraphics[width=0.7\linewidth]{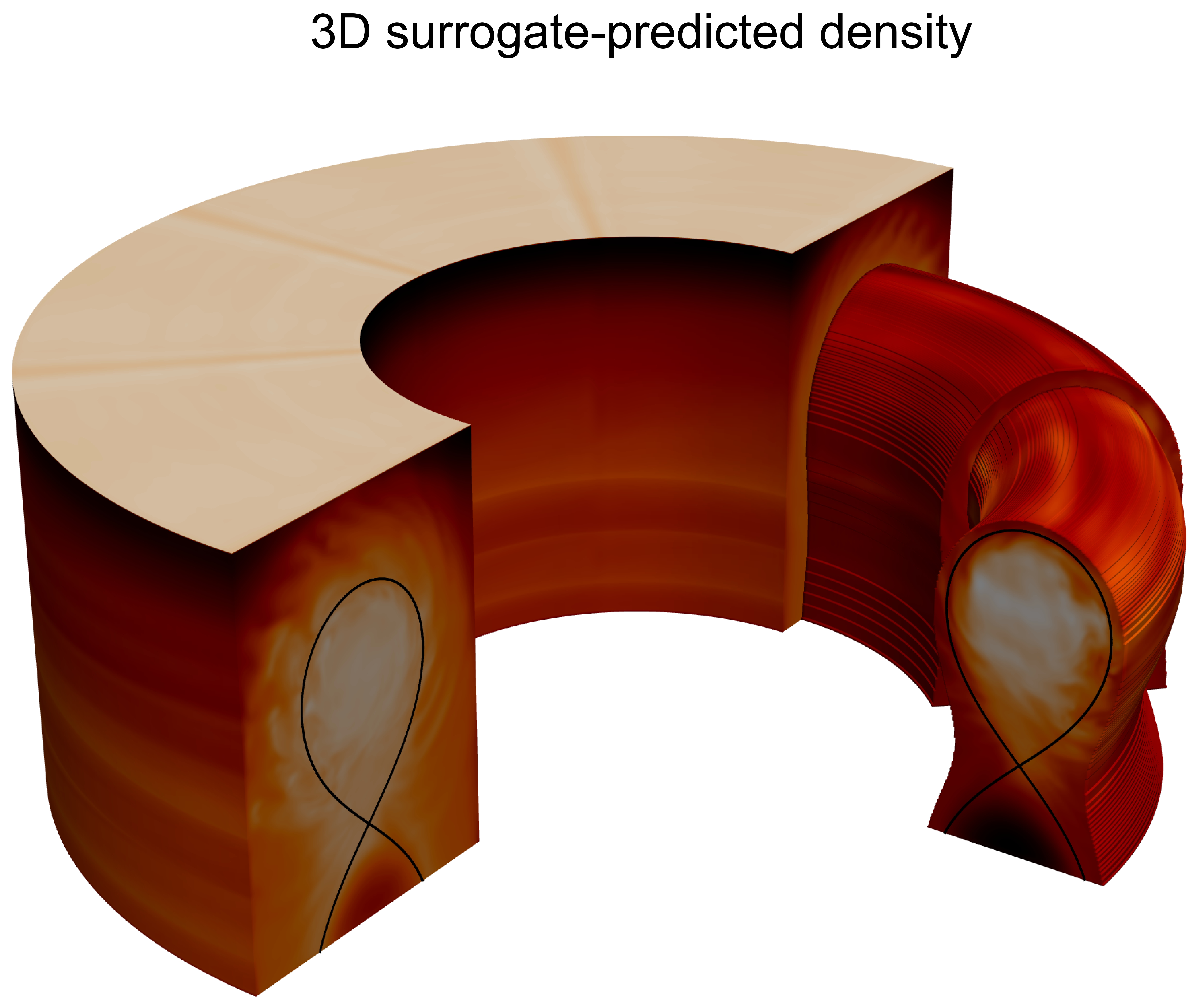}
\caption{Three-dimensional cutaway of the toroidal simulation domain at held-out $\nu_0=0.1$, colored by the surrogate-predicted electron density $n=\exp(\theta)$ produced by the resistivity-conditioned KNO. The high-density core transitions through a steep gradient to the low-density scrape-off layer, with magnetic field lines overlaid on the poloidal cross-sections.}
\label{fig:torus_3d}
\end{figure}

The slice metrics above quantify local one-step fidelity. Figure~\ref{fig:torus_3d} complements this analysis by showing the surrogate-predicted electron density, $n=\exp(\theta)$, over the retained three-dimensional toroidal subset. The prediction preserves a coherent toroidal density structure, including the transition from the high-density core to the lower-density SOL and visible turbulent structures in the boundary region. Although this figure is intended as a qualitative visualization rather than a full volumetric error analysis, it indicates that the KNO output remains spatially coherent in three dimensions. Such low-cost three-dimensional predictions may be useful as approximate initial states for GBS restarts, reducing the need to evolve full simulations through the initial growth phase before reaching quasi-steady turbulence~\cite{Solheim2026}.

\subsection{Pressure-gradient length}
\begin{figure}[H]
\centering
\includegraphics[width=\linewidth]{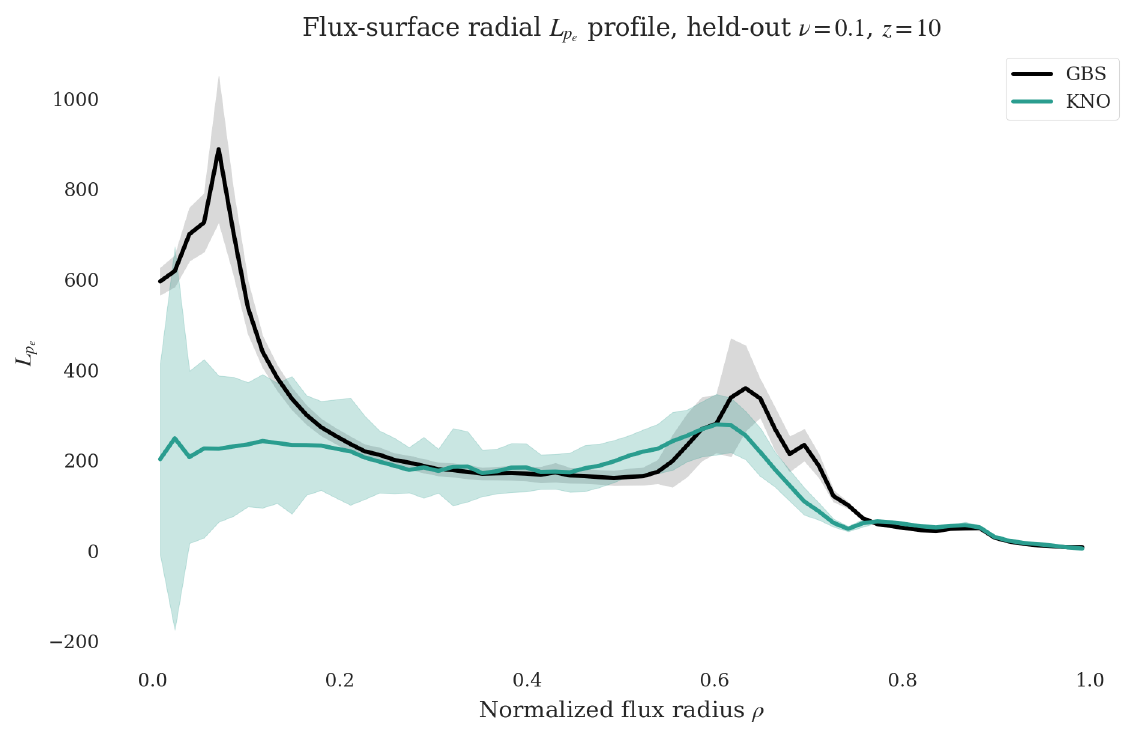}
\caption{Radial profiles of the time-averaged electron pressure-gradient scale length $L_{p_e}$ at held-out $\nu_0=0.1$: GBS reference and KNO prediction. Shading indicates variability where provided by the plotting routine. The average is taken over poloidal angle and 50 snapshots with timestep $\Delta t=0.1$ on the poloidal slice at toroidal plane index $i_\varphi=10$.}
\label{fig:Lpe}
\end{figure}

Beyond instantaneous field agreement, a physically meaningful test is whether the surrogate reproduces reduced radial quantities relevant to turbulence drive. We assess this using the electron pressure-gradient scale length $L_{p_e}=-p_e/|\nabla p_e|$, which characterizes the local pressure-gradient strength available to drive interchange instabilities and is closely related to radial SOL-width scaling. Figure~\ref{fig:Lpe} compares the time-averaged radial profile $L_{p_e}$ between the GBS reference and the KNO prediction, averaged over poloidal angle and 50 snapshots with $\Delta t=0.1$.

The surrogate captures the broad mid-radius and outer-region structure of $L_{p_e}$, indicating that the radial pressure-gradient structure is statistically reproduced at the held-out resistivity. The sharpest inner-gradient features are less accurately recovered, consistent with the tendency of finite-bandwidth spectral operators to smooth localized gradients. This result shows that the KNO does not merely interpolate pointwise field values, but also preserves an integrated boundary-plasma diagnostic relevant to turbulence drive.

\subsection{Teacher-forced temporal statistics}\label{sec:val:autocorr}

\begin{figure}[H]
\centering
\includegraphics[width=\linewidth]{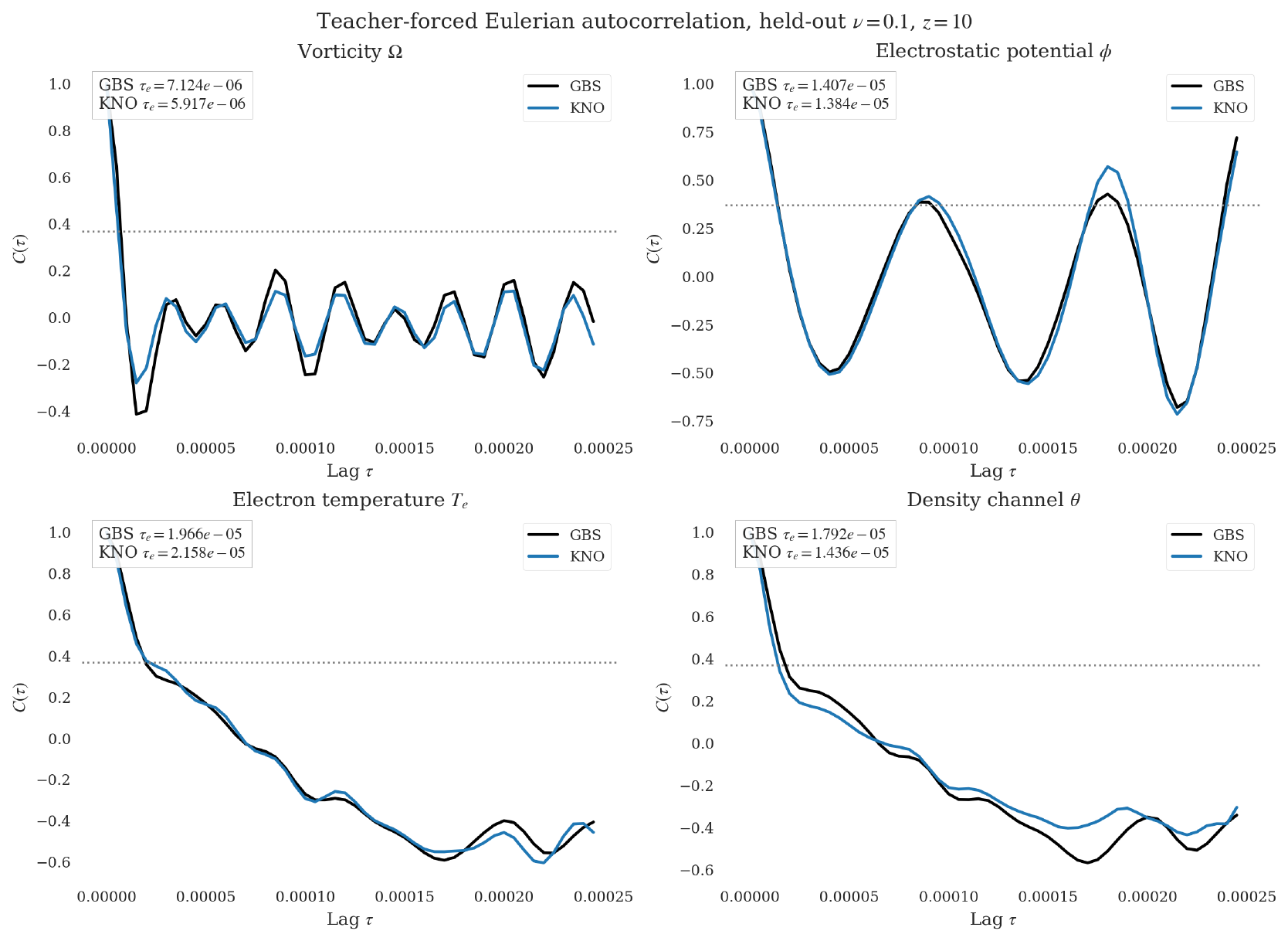}
\caption{Teacher-forced Eulerian autocorrelation $C(\tau)$ at the held-out resistivity $\nu_0=0.1$ for $\Omega$, $\phi$, $T_e$, and $\theta$. GBS reference curves are shown in black and KNO predictions in blue. Annotated e-folding times $\tau_e$ are shown where the plotting routine converges. Fluctuations are formed by subtracting the per-pixel time mean before computing the autocorrelation.}
\label{fig:autocorr_tf}
\end{figure}

We next examine whether the surrogate preserves short-time temporal memory when it is kept close to the reference trajectory. Figure~\ref{fig:autocorr_tf} compares Eulerian autocorrelations at held-out $\nu_0=0.1$ under teacher forcing. In this diagnostic, the input history is taken from the GBS trajectory at each step, so the test isolates conditional temporal statistics from error accumulation during free-running rollout.

The KNO reproduces the GBS autocorrelation well for $\phi$, $T_e$, and $\theta$, showing that the model captures the short-time temporal decorrelation of these fields when it is conditioned on reference histories. For $\Omega$, the surrogate also captures the rapid loss of correlation and the remaining oscillatory behavior with comparable e-folding times, although vorticity remains the most challenging field in terms of instantaneous slice accuracy. Overall, the teacher-forced statistics indicate that the model has learned meaningful short-horizon conditional dynamics. However, they do not by themselves demonstrate that the same dynamics remain stable during fully autoregressive rollouts, where no reference correction is applied.

\subsection{Spectral fidelity and scale-by-scale errors}\label{sec:val:spectra}

\begin{figure}[H]
\centering
\includegraphics[width=\linewidth]{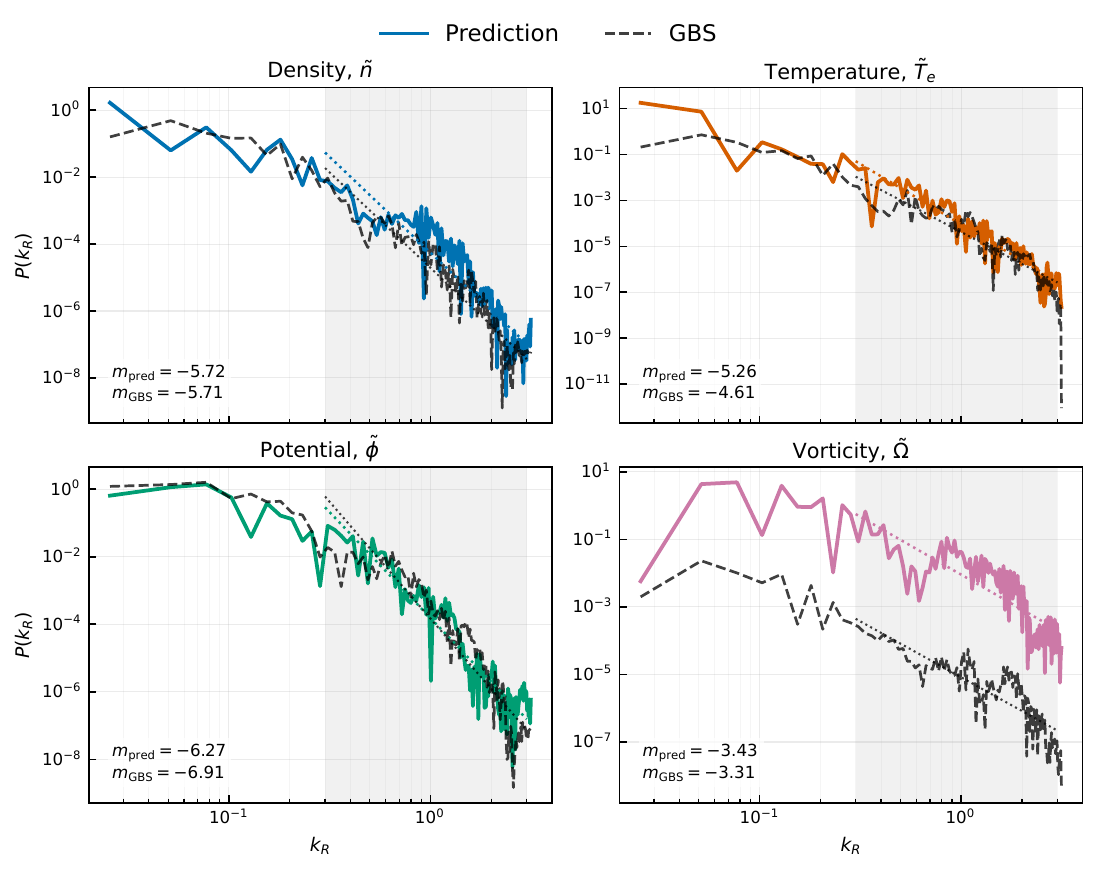}
\caption{Radial power spectra (log--log) at held-out $\nu_0 = 0.1$, one-step prediction: KNO (solid) versus GBS (dashed) for $\tilde{n}$, $\tilde{T}_e$, $\tilde{\phi}$, and
$\tilde{\Omega}$. Tildes ($\tilde{.}$) denote fluctuating components about the mean, $\tilde{A} = A - \langle A \rangle$. Dotted lines are log--log slope fits ($m_\mathrm{pred}$, $m_\mathrm{GBS}$). }
\label{fig:radial_spectra}
\end{figure}

We now turn from pointwise and reduced-profile diagnostics to scale-by-scale turbulent structure. Spectral fidelity is assessed using radial power spectra on a midplane $k_R$ band. Figure~\ref{fig:radial_spectra} compares the one-step KNO and GBS spectra, while Table~\ref{tab:fieldwise-summary} reports the corresponding log--log slope fits. Slopes are fit over $0.3\le k_R\le 3.0$.

At the held-out value $\nu_0 = 0.1$, the density channel gives the closest spectral-slope agreement, with a relative slope error of $0.2\%$. The potential
and temperature channels have slope errors of $9.3\%$ and $14.1\%$ respectively, with the temperature surrogate overestimating large-scale ($k_R \lesssim 0.3$) power by one to two orders of magnitude before recovering the correct spectral trend at mid-to-high wavenumbers.

Vorticity shows the largest spectral discrepancy. Although its fitted slope error is modest, $3.6\%$ with $m_\mathrm{pred}=-3.43$ and $m_\mathrm{GBS}=-3.31$, the KNO overestimates broadband vorticity power by roughly two to three orders of magnitude across the resolved $k_R$ range. The two spectra are nearly parallel in log--log space but vertically offset, indicating that the surrogate captures the spectral exponent but not the absolute fluctuation level. This is the spectral signature of the unenforced polarization coupling discussed above: with the amplitude relation between $\Omega$ and $\phi$ left unconstrained, the vorticity spectrum can retain the correct slope while settling at an incorrect level. The slope error alone therefore overstates vorticity fidelity, and broadband amplitude agreement should be treated as a separate diagnostic requirement.

\begin{figure}[H]
\centering
\includegraphics[width=\linewidth]{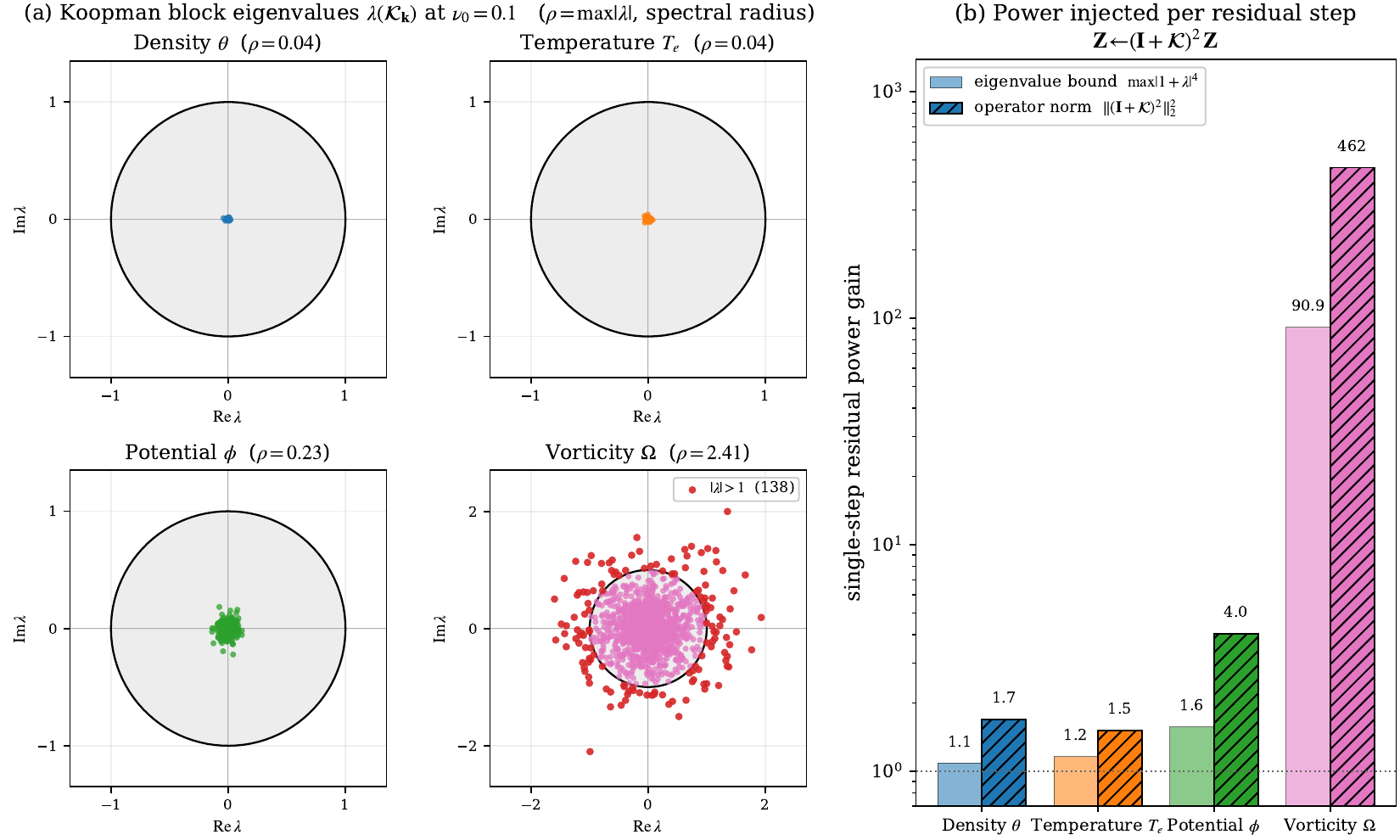}
\caption{Spectral stability of the learned Koopman operators at the held-out $\nu_0=0.1$. \textbf{(a)} Eigenvalues $\lambda(\mathcal{K}_{\mathbf{k}})$ of the per-mode block of the spectral operator $\mathcal{K}$ on the complex plane, against the unit circle: only vorticity leaves the unit disk (spectral radius $\rho_\Omega=2.41$, with $138$ unstable eigenvalues), whereas $\theta$, $T_e$, and $\phi$ remain tightly contractive ($\rho\le0.23$; each field in its own panel). \textbf{(b)} Single-step power gain of the residual update $\mathbf{Z}\leftarrow(\mathbf{I}+\mathcal{K})^2\mathbf{Z}$: the eigenvalue bound $\max|1+\lambda|^{4}$ and the operator norm $\lVert(\mathbf{I}+\mathcal{K})^2\rVert_2^{2}$.}
\label{fig:koopman_stability}
\end{figure}

Vorticity's spectral amplitude overestimate and rollout divergence both arise from a single non-contractive learned operator. The vorticity operator has spectral radius $\rho_\Omega=2.41$ (eigenvalues outside the unit circle), so it amplifies fluctuations at every step, whereas the operators for the other three fields are contractive ($\rho\le0.23$). The $\tanh$ output activation masks this instability in one-step prediction by capping the variance, but the amplification compounds under autoregressive rollout, contributing to the error growth in \cref{fig:rollout_l2}. The learned vorticity update has a single-step power gain of $90.9\times$ (eigenvalue bound) to $462\times$ (operator norm), versus at most $4\times$ for the other three fields (\cref{fig:koopman_stability}). This isolated non-contractivity is the operator-level counterpart of the same polarization coupling~\eqref{GBS_Poisson}, which the fieldwise surrogate does not enforce.

\subsection{Blob detection}\label{sec:val:blobs}

\begin{figure}[H]
\centering
\includegraphics[width=\linewidth]{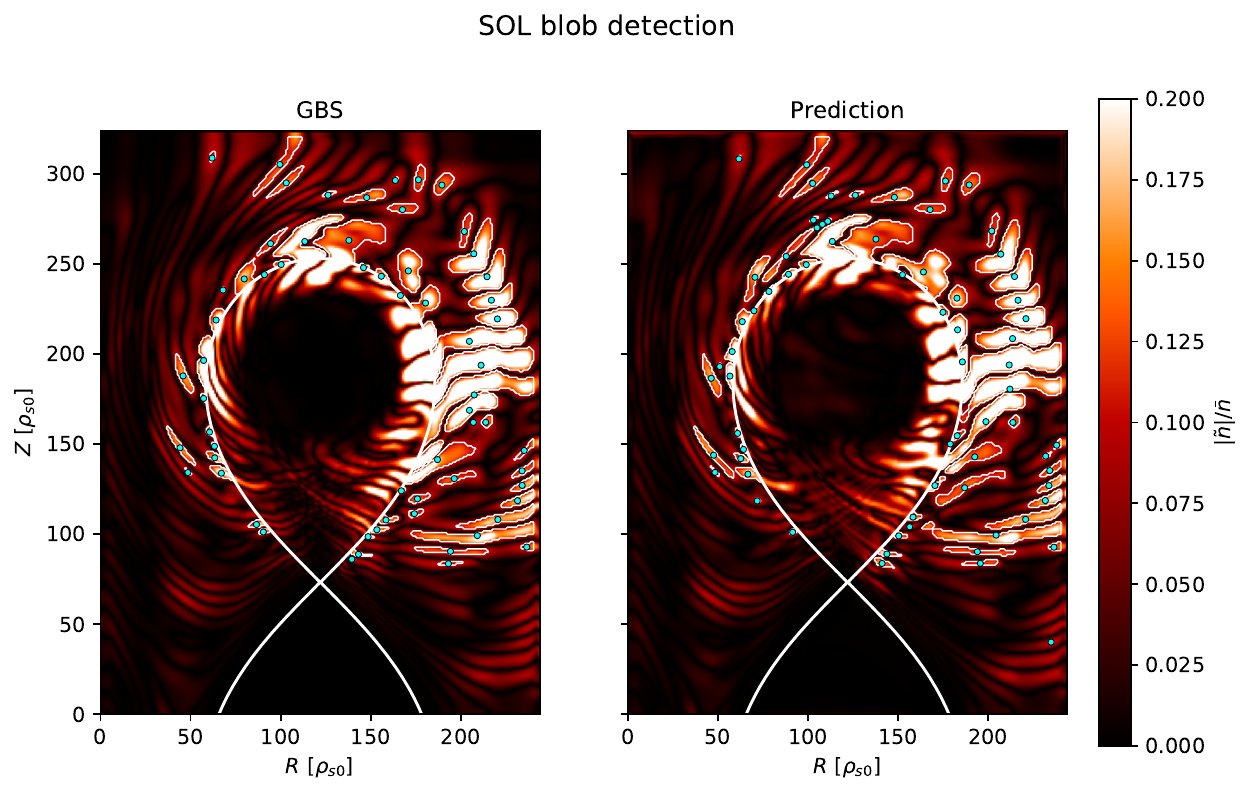}
\caption{Normalized density fluctuation $|\tilde{n}|/\bar{n}$ on the poloidal $(R,Z)$ plane at the held-out resistivity $\nu_0=0.1$. White contours and cyan centroids indicate blobs detected using the threshold $|\tilde{n}|/\bar{n}>0.10$ in the SOL region where the background density $\bar{n}$ (toroidal average over the 20 retained planes) is taken from the GBS reference for \emph{both} fields to remove normalization differences. The surrogate captures the blob structures and radial locations well.}
\label{fig:blob_detection_2d}
\end{figure}

The spectral diagnostics identify how the surrogate distributes fluctuation power across scales. We next examine how these scale-dependent errors appear in physical space through blob detection technique. Figure~\ref{fig:blob_detection_2d} overlays blob-detection contours on the normalized density fluctuation $|\tilde{n}|/\bar{n}$ in the poloidal plane at the held-out resistivity $\nu_0=0.1$, using a threshold $|\tilde{n}|/\bar{n}>0.10$ in the SOL region following Ref.~\cite{Lim2026} with a common GBS-derived background $\bar{n}$ (toroidal average over the 20 retained planes at the matched one-step frame).

The KNO prediction captures the main SOL blob structures and their approximate radial locations, including elongated features along the separatrix and in the outer SOL. With the common GBS background, the one-step blob populations agree quantitatively (63 versus 71 blobs; blob-mask IoU 0.79). This is the physical-space counterpart of the small-scale spectral over-prediction in \cref{sec:val:spectra}. 

\subsection{Autoregressive rollout and stability limit}\label{sec:val:rollout}

The diagnostics discussed above are based on either one-step prediction or teacher-forced evaluation. They therefore measure the conditional statistical fidelity of the surrogate while its input remains close to the reference GBS trajectory. To assess autonomous performance, we perform autoregressive rollouts for up to 12 steps at the held-out value $\nu_0=0.1$, where each predicted field is fed back into the input history for the next prediction. The corresponding one-step persistence baselines are reported in Table~\ref{tab:fieldwise-summary}.
 
\begin{figure}[H]
\centering
\includegraphics[width=\linewidth]{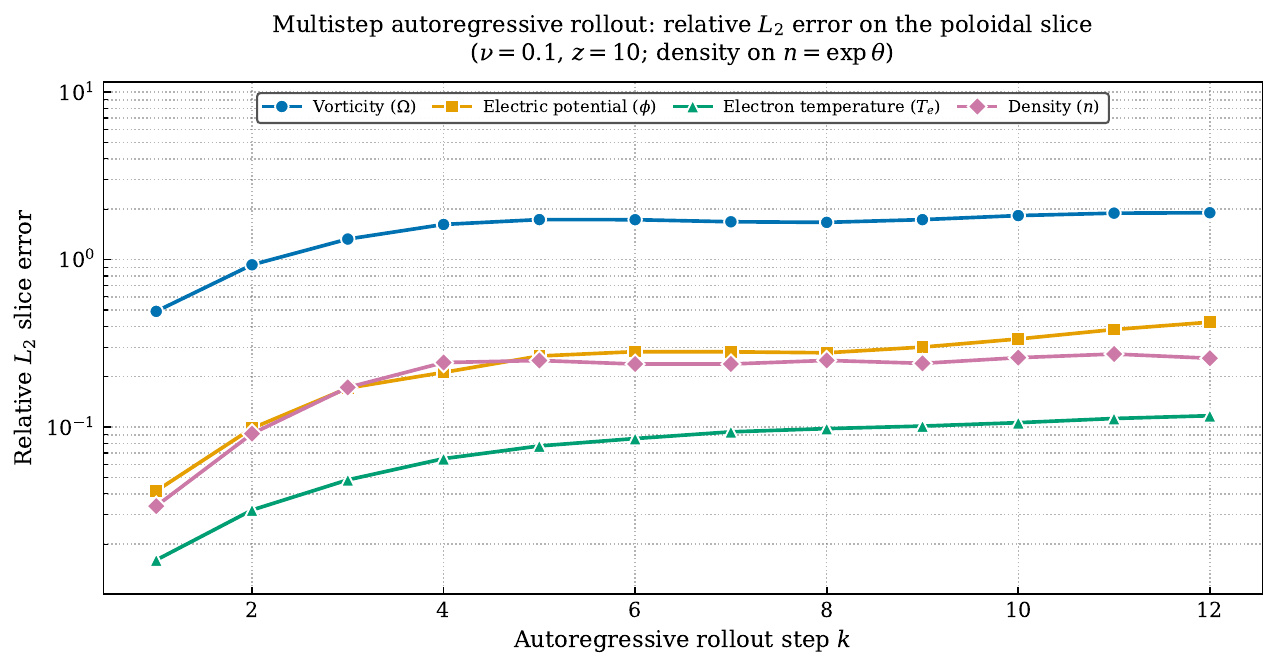}
\caption{Relative $L_2$ rollout error over the first 12 autoregressive steps at the held-out $\nu_0=0.1$ on the poloidal slice $i_\varphi=10$, for $\Omega$, $\phi$, $T_e$, and density ($n=\exp(\theta)$). Vorticity grows fastest, consistent with its non-contractive learned operator (\cref{fig:koopman_stability}), while density, potential, and temperature remain comparatively stable over this window. The vertical axis is logarithmic.}
\label{fig:rollout_l2}
\end{figure}

Figure~\ref{fig:rollout_l2} shows that good one-step accuracy does not necessarily lead to stable free-running predictions. In training and teacher-forced validation, each surrogate is conditioned on ground-truth histories over a short prediction window. In autoregressive rollout, however, the model is fed with its own previous predictions. As soon as the predicted state moves away from the distribution seen during ground-truth-conditioned prediction, errors can accumulate and the trajectory may drift from the GBS reference.

The 12-step rollout window highlights the field-dependent divergence rate: vorticity accumulates error fastest, in line with its non-contractive Koopman operator, whereas density, potential, and electron temperature grow more slowly while the input stays near the reference trajectory. Overall, the rollout results define the present scope of the model: an effective short-horizon, resistivity-conditioned statistical emulator.

\subsection{Computational cost}\label{sec:val:cost}
Finally, we report the inference cost of the retained KNO configuration. Table~\ref{tab:archbench} lists the trainable parameter count and forward-pass latency. Since all four fieldwise models use the same Fourier-mode set, their parameter counts are identical. A direct wall-clock comparison with GBS would require matched hardware, solver settings, grid resolution, timestep, and output cadence, which can vary substantially between runs. We therefore report the neural inference latency directly, rather than quoting a universal speedup factor. A full drift-reduced Braginskii time integration remains substantially more expensive than a single neural forward pass~\cite{Dudson2009,Chang2009}. The measured per-field forward time of approximately ${124}$\,ms indicates that the resistivity-conditioned KNO is suitable for rapid exploratory studies and diagnostic-dashboard applications within the parameter range considered here.

\begin{table}[H]
\centering
\setlength{\tabcolsep}{3pt}
\begin{tabular}{@{}p{0.60\linewidth}p{0.18\linewidth}@{}}
\hline
Quantity & Value \\
\hline
Trainable parameters ($\phi$, $T_e$, $\theta$; each) & $\trainparams$ \\
Trainable parameters (vorticity $\Omega$) & $\trainparamsomega$ \\
Mean forward time & $\sim\inferms$~ms \\
Field-to-latent ratio ($n_{\mathrm{fields}}/d_\ell$, $n_{\mathrm{fields}}=8$) & $\sim\channelratio$ \\
\hline
\end{tabular}
\caption{Architecture benchmark for the retained $\nu_0$-conditioned fieldwise configuration. Parameter counts are exact trainable totals. Forward time is measured for the retained electron-temperature surrogate at batch size one on an NVIDIA A100 80GB PCIe using PyTorch 2.5.1+cu121, giving $123.64\pm0.65$~ms over 200 timed passes, with a minimum of $122.14$~ms.}
\label{tab:archbench}
\end{table}

These numbers characterize the neural inference cost for the retained configuration. The computational advantage should therefore be interpreted together with the validation results above: the KNO is inexpensive and useful for short-horizon interpolation and diagnostic evaluation, while stable long-horizon replacement of the GBS dynamics remains an open challenge.

%% file: results/fieldwise_summary_table.tex
\begin{table}[H]
    \centering
    \footnotesize
    \setlength{\tabcolsep}{4pt}
    \begin{tabular}{@{}lcccccccc@{}}
    \hline
    Field & MSE & $R^2$ & $\|\hat u-u\|_2/\|u\|_2$ & $m_{\mathrm{GBS}}$ & $m_{\mathrm{pred}}$ & $\dfrac{m_{\mathrm{pred}}-m_{\mathrm{GBS}}}{|m_{\mathrm{GBS}}|}$ & KNO $\|\cdot\|_2$ & Pers.\ $\|\cdot\|_2$ \\
    \hline
        $\Omega$ & $7.76\times 10^{-5}$ & 0.759 & 0.491 & $-3.31$ & $-3.43$ & $-0.036$ & 0.491 & 0.784 \\
        $\phi$ & $1.20\times 10^{-2}$ & 0.998 & 0.0416 & $-6.91$ & $-6.27$ & 0.093 & 0.0416 & 0.0798 \\
        $T_{\mathrm{e}}$ & $1.70\times 10^{-4}$ & 0.9997 & 0.0161 & $-4.61$ & $-5.26$ & $-0.141$ & 0.0161 & 0.0329 \\
        $n$ & $9.94\times 10^{-4}$ & 0.9982 & 0.0338 & $-5.71$ & $-5.72$ & $-0.002$ & 0.0338 & 0.0487 \\
    \hline
    \end{tabular}
    \caption{Fieldwise one-step metrics for the held-out case $\nu_0=0.1$ at $i_\varphi=10$ within the retained 20-plane subset. MSE, $R^2$, and relative $L_2$ quantify slice accuracy; $m_{\mathrm{GBS}}$ and $m_{\mathrm{pred}}$ are midplane $k_R$ spectral slopes, with relative slope error measuring spectral mismatch. The last two columns compare KNO and persistence relative $L_2$ errors.}
    \label{tab:fieldwise-summary}
\end{table}

%% file: sections/5_Conclusion.tex
\section{Conclusion}\label{sec:conclusion}
We developed resistivity-conditioned KNOs for electrostatic drift-reduced Braginskii turbulence using three-dimensional, flux-driven, nonlinear GBS simulations. The nominal resistivity label $\nu_0$ was used as the conditioning parameter because it controls the resistive response and separates boundary-plasma regimes with different profile relaxation, fluctuation activity, and spectral content. Independent fieldwise surrogates were trained for the learned density channel $\theta=\ln n$, electron temperature $T_e$, electric potential $\phi$, and vorticity $\Omega$. Their performance was assessed at the held-out resistivity $\nu_0=0.1$, providing an in-range interpolation test.

The key positive result is that the model achieves good statistical fidelity at an unseen resistivity over short horizons---the one-step forecast interval $\Delta t_{\mathrm{out}}=0.1\,(R_0/c_{s0})$ on which the pointwise, spectral, and teacher-forced diagnostics are evaluated, with free-running rollouts assessed up to 12 steps ($\approx 1.2$). The $T_e$ surrogate achieves near-unity one-step slice agreement ($R^2>0.999$), while the density and electric-potential surrogates remain highly accurate ($R^2>0.99$). Among the four fields, the density channel preserves the radial spectral slope most accurately. Reduced physical diagnostics, such as the time-averaged electron pressure-gradient scale length, are also well reproduced. The density-fluctuation analysis further shows that the surrogate captures the main SOL blob locations and gross morphology, although smaller-scale structures and local amplitudes are less accurately recovered. These results indicate that resistivity-conditioned KNOs can provide fast conditional emulators for selected pointwise, spectral, and reduced physical diagnostics within the parameter range considered.

The main limitations are field dependence and free-running stability. Among the four fieldwise surrogates, $\Omega$ remains the most challenging field, because it is a derivative quantity, is sensitive to high-wavenumber fluctuations, and is predicted independently from $\phi$ without explicitly enforcing the polarization relation. In addition, the autoregressive rollouts progressively depart from the GBS reference trajectory, despite good one-step and teacher-forced performance. The present model therefore captures useful short-horizon conditional statistics, but does not yet provide a stable long-horizon closure for the full GBS dynamics.

Overall, these results support the use of resistivity-conditioned fieldwise neural operators as fast surrogates for rapid exploratory studies, interpolation tests, and diagnostic-dashboard applications, especially when repeated full GBS simulations are costly. Stable long-horizon emulation and coupled multi-field consistency remain open challenges. Future work should therefore focus on coupled-field training, physics-informed constraints on the vorticity--potential relation, rollout-aware training objectives, and nonlinear latent dynamics that can better represent turbulent mixing beyond short-step Koopman evolution; a concrete instance is constraining the vorticity operator to be contractive (spectral radius below unity), as it already is for the other three fields, which would suppress the spurious small-scale power amplification identified here.

%% file: sections/8_Appendix.tex
\section{Implementation details}\label{sec:appendix-impl}

This appendix records implementation details that complement the architecture summarised in \cref{sec:kno_framework}. Equations already given there---the temporal lift and spatial encoding (\cref{eq:temporal_encoder,eq:3D}), the additive resistivity-conditioned operator (\cref{eq:Kd_additive}), the residual latent update (\cref{eq:kno_residual_update}), the skip and decode (\cref{eq:skip,eq:kno_decode}), and the training objective (\cref{eq:Ltotal,eq:Lpred})---are referenced rather than repeated.

\subsection{Spectral operator: explicit form}\label{sec:spectral-appendix}

After the spatial encoding, the latent field is transformed to Fourier space and updated by the direction-wise complex matrices of \cref{eq:Kd_additive}. Writing $\tilde{\mathbf{Z}}$ for the transformed latent, the update is applied sequentially across the retained directions,
\begin{align}
\tilde{\mathbf{Z}}^{[R]}(\mathbf{k}) &= \sum_{j=1}^{\latentdim} \mathcal{K}_{R}^{ij}(\nu_0,\Delta t,k_R)\,\tilde{\mathbf{Z}}_j(\mathbf{k}), \\
\tilde{\mathbf{Z}}^{[Z]}(\mathbf{k}) &= \sum_{j=1}^{\latentdim} \mathcal{K}_{Z}^{ij}(\nu_0,\Delta t,k_Z)\,\tilde{\mathbf{Z}}^{[R]}_j(\mathbf{k}), \\
\tilde{\mathbf{Z}}^{[\varphi]}(\mathbf{k}) &= \sum_{j=1}^{\latentdim} \mathcal{K}_{\varphi}^{ij}(\nu_0,\Delta t,k_\varphi)\,\tilde{\mathbf{Z}}^{[Z]}_j(\mathbf{k}).
\end{align}
We write $\mathcal{K}(\mathbf{Z},\nu_0,\Delta t)$ for the operator on grid-valued latents obtained by composing the forward transform, the three directional updates above on the truncated mode cube, and the inverse transform back to real space. At a single Fourier triplet $\mathbf{k}=(k_R,k_Z,k_\varphi)$ the action of $\mathcal{K}$ reduces to a complex block $\mathcal{K}_{\mathbf{k}}=\mathcal{K}_{\varphi}(k_\varphi)\,\mathcal{K}_{Z}(k_Z)\,\mathcal{K}_{R}(k_R)\in\mathbb{C}^{d_\ell\times d_\ell}$ acting on the $d_\ell=\latentdim$ latent channels. Only a truncated set of $\modesx\times\modesy\times\modesz$ Fourier modes is retained, so the spectral block acts as a practical low-mode regulariser, distinct from the over-complete latent width $d_\ell=\latentdim$.

\subsection{Residual evolution and decoding}\label{sec:residual-appendix}

The composed operator enters the residual latent update of \cref{eq:kno_residual_update}, applied for $N_{\mathrm{iter}}=\repetitions$ steps, after which the skip combination and temporal decoder (\cref{eq:skip,eq:kno_decode}) emit the physical prediction. No intermediate nonlinearity is applied between the $N_{\mathrm{iter}}$ residual updates; the pointwise $\tanh$ appears only in the temporal encoder and in the final skip. The residual form supplies a short-time correction to the latent state rather than replacing it: during training the temporal decoder jointly emits $\predsteps$ frames, while the one-step slice metrics of \cref{sec:validation} use only the first.

\subsection{Spectral stability of the learned operators}\label{sec:spectral-stability-appendix}

For each fieldwise surrogate we extract, from the retained checkpoint at the held-out resistivity $\nu_0=0.1$, the per-mode block $\mathcal{K}_{\mathbf{k}}$ of the composed spectral operator $\mathcal{K}$ (\cref{sec:spectral-appendix}) at each Fourier triplet $\mathbf{k}=(k_R,k_Z,k_\varphi)$. Because the latent state is advanced by the residual update $\mathbf{Z}\leftarrow\mathbf{Z}+\mathcal{K}(\mathbf{Z})$ applied $N_{\mathrm{iter}}=\repetitions$ times per step (\cref{eq:kno_residual_update}), the single-step amplification of modal power is bounded below by the eigenvalue gain $\max_i|1+\lambda_i(\mathcal{K}_{\mathbf{k}})|^{2N_{\mathrm{iter}}}$ and exactly by the operator-norm gain $\lVert(\mathbf{I}+\mathcal{K}_{\mathbf{k}})^{N_{\mathrm{iter}}}\rVert_2^{2}$; the two coincide for a normal operator and diverge under non-normal transient growth. We evaluate both over the $4\times4\times4=64$ lowest-order Fourier modes (results are insensitive to including higher modes, which are strongly contractive for all fields). The resulting per-field gains are reported in \cref{fig:koopman_stability}.

\subsection{Network blocks}

The surrogate contains four main components.
\begin{itemize}
\item A temporal encoder-decoder pair operating pointwise at each grid point. The encoder is a single learned linear layer (bias included) from the $H$-history to a latent vector of width \latentdim, followed by a pointwise $\tanh$ activation. The decoder is a single learned linear layer from width \latentdim\ to a vector of width $\predsteps$ (the joint prediction horizon) with no activation. The decoder width matches the prediction horizon rather than the input history, so the temporal encoder-decoder round trip functions as an auxiliary prediction head rather than a strict reconstruction.
\item A spatial encoder-decoder pair built from three-dimensional convolutions that mix neighbouring structure on the simulation grid. Each of the encoder and decoder is a stack of three $\mathrm{Conv3d}$ layers with ReLU activations; channel widths are $(\latentdim, 4{\times}\latentdim, \latentdim)$ inside the stack so that the spatial tensor entering and leaving the stack has width \latentdim.
\item A direction-wise Fourier-space update whose learned matrices depend on the conditioning parameter $\nu_0$ and the timestep $\Delta t$. In minibatches that contain more than one $\nu_0$ (or $\Delta t$) value, the complex corrections $\mathcal{K}_d^{(\nu)}(\nu_0)$ and $\mathcal{K}_d^{(\Delta t)}(\Delta t)$ are averaged across the batch before application, so the effective operator within a batch is the scalar-averaged version. The implementation applies a truncated complex Fourier cube on $(R,Z,\varphi)$ with axis ordering and \texttt{torch.fft.rfftn} conventions matching the released code.
\item A final skip $\mathbf{z}^{\mathrm{final}}=\tanh(\mathbf{W}_0\star\mathbf{z}+\mathcal{D}_\mathrm{3D}(\mathbf{Z}^{(N_\mathrm{iter})}))$, where $\mathbf{W}_0$ is a $1\times1\times1$ $\mathrm{Conv3d}$ preserving the latent-channel width $\latentdim$. This skip sits outside the $\mathcal{E}_\mathrm{3D}/\mathcal{D}_\mathrm{3D}$ pair, so no additional $\mathcal{E}_\mathrm{3D}$ pass is applied on that branch.
\end{itemize}

Mapping from paper symbols to code field identifiers used in the released repository: $\Omega\leftrightarrow\texttt{omega}$, $\phi\leftrightarrow\texttt{strmf}$ (the GBS streamfunction label denotes the same electrostatic potential $\phi$), $T_e\leftrightarrow\texttt{temperature}$, and the learned density channel $\theta\leftrightarrow\texttt{theta}$ (physical density $n=\exp(\theta)$ when mapped). All four fieldwise models retain up to $\modesx\times\modesy\times\modesz$ Fourier modes and apply $\repetitions$ residual updates in latent space.

\subsection{Auxiliary prediction and latent-consistency losses}

The auxiliary stack $\mathcal{L}_{\mathrm{aux}}$ of \cref{eq:Ltotal} contains one auxiliary prediction term and three latent-consistency terms evaluated in MSE,
\begin{align}
\mathcal{L}_{\mathrm{aux}} &= \mathrm{MSE}\bigl(\mathcal{D}_\mathrm{temp}(\mathbf{z}),\,\mathbf{f}^{\mathrm{target}}_{1:\predsteps}\bigr)
+ \mathrm{MSE}\bigl(\mathcal{E}_\mathrm{temp}(\mathcal{D}_\mathrm{temp}(\mathbf{z})),\,\mathbf{z}\bigr) \nonumber\\
&\quad + \mathrm{MSE}\bigl(\mathcal{D}_{\mathrm{3D}}(\mathcal{E}_{\mathrm{3D}}(\mathbf{Z}_{\mathrm{pre}})),\,\mathbf{Z}_{\mathrm{pre}}\bigr)
+ \mathrm{MSE}\bigl(\mathcal{E}_{\mathrm{3D}}(\mathcal{D}_{\mathrm{3D}}(\mathbf{Z}_{\mathrm{post}})),\,\mathbf{Z}_{\mathrm{post}}\bigr),
\end{align}
where $\mathbf{z}=\tanh(\mathcal{E}_\mathrm{temp}(\cdot))$ is the temporal latent of \eqref{eq:temporal_encoder}, $\mathbf{Z}_{\mathrm{pre}}=\mathbf{Z}^{(0)}=\mathcal{E}_\mathrm{3D}(\mathbf{z})$ is the spatial latent before the Koopman stack, and $\mathbf{Z}_{\mathrm{post}}=\mathbf{Z}^{(N_\mathrm{iter})}$ is the latent after $N_\mathrm{iter}$ residual updates. Because the temporal decoder has output width $\predsteps$, the first term is an auxiliary four-frame prediction head trained against $\mathbf{f}^{\mathrm{target}}_{1:\predsteps}$ rather than a reconstruction of the input history. The remaining three terms regularise temporal and spatial encode-decode consistency before and after the Koopman stack; they are practical regularisers that encourage informative latents and stable decoding, not proofs of invertibility or physical consistency.

\subsection{Retained checkpoint optimisation metrics}

\Cref{tab:fieldwise-train-appendix} lists the training epoch and composite optimisation objectives for the same fieldwise checkpoints summarised in \cref{tab:fieldwise-summary}. \emph{Train} and \emph{Val} are the main prediction losses (components of $\mathcal{L}_{\mathrm{pred}}$, \cref{eq:Lpred}) on the training and within-trajectory validation loaders at the listed epoch; \emph{Val pred}, \emph{Val recon}, and \emph{Train pred} are auxiliary-stack components introduced alongside \cref{eq:Ltotal}. These entries are not physical error percentages.

\begin{table}[htbp]
\centering
\footnotesize
\setlength{\tabcolsep}{4pt}
\begin{tabular}{@{}lrrrrrr@{}}
\hline
Field & Epoch & Train & Val & Val pred & Val recon & Train pred \\
\hline
$\Omega$ & 19980 & 0.012206 & 0.012008 & 0.012008 & 1.799853 & 0.012206 \\
$\phi$ & 17720 & 0.000733 & 0.000732 & 0.000732 & 0.062821 & 0.000733 \\
$T_{\mathrm{e}}$ & 13180 & 0.000565 & 0.000565 & 0.000565 & 0.007843 & 0.000565 \\
$\theta$ & 19340 & 0.000404 & 0.000422 & 0.000422 & 0.005157 & 0.000404 \\
\hline
\end{tabular}
\caption{Retained checkpoint epochs and composite losses for the fieldwise runs referenced in \cref{tab:fieldwise-summary}.}
\label{tab:fieldwise-train-appendix}
\end{table}

\subsection{Latent width and memory}

The latent tensor uses $d_\ell=\latentdim$ scalar channels per grid point, which is wider than the eight physical fields of the full Braginskii state (expansion ratio $d_\ell/n_{\mathrm{fields}}\approx\;4$). This over-complete representation provides additional degrees of freedom for the Koopman-style spectral update to approximate nonlinear dynamics in a lifted, approximately linear space. Storing latent activations during training therefore scales with $d_\ell$ rather than implying channel-wise reduction relative to the full multi-field state.

\section{Toroidal dimension reconstruction}\label{sec:z-recon}

The surrogate is trained on a contiguous subset of \gridzsubset\ toroidal planes (indices $0,\ldots,\gridzsubset-1$ of the full \gridz-plane GBS grid). Because the toroidal coordinate $\varphi$ is $2\pi$-periodic, one can in principle recover the full toroidal field from a reduced set of samples by spectral reconstruction. This section records the reconstruction method and its conditioning.

\subsection{GBS toroidal grid convention}

The GBS simulation stores \gridz\ toroidal planes at coordinates
\begin{equation}
 \varphi_k = k\,\frac{2\pi}{N_\varphi}, \qquad k=0,\ldots,N_\varphi-1, \quad N_\varphi=\gridz,
\end{equation}
i.e.\ an endpoint-exclusive periodic partition with uniform spacing $\Delta\varphi=2\pi/N_\varphi$. The reconstruction target grid must match this convention exactly; an endpoint-inclusive grid (spacing $2\pi/(N_\varphi-1)$) introduces a coordinate bias that grows with toroidal index and corrupts the reconstructed field.

\subsection{Complex-exponential Fourier reconstruction}

Given field values $f(\varphi_{j})$ at $N_s=\gridzsubset$ sparse toroidal locations, the reconstructor fits a truncated complex Fourier series using the basis $\{e^{ik\varphi}\}_{k=-K}^{K}$ with $K=10$, giving $2K+1=21$ complex basis functions. The design matrix $\mathbf{B}_s \in \mathbb{C}^{N_s \times (2K+1)}$ is assembled by evaluating $e^{ik\varphi_j}$ at each sample location, yielding a slightly underdetermined complex system ($N_s=20 < 2K+1=21$). The coefficient vector is found as the minimum-norm least-squares solution,
\begin{equation}
 \mathbf{c} = \arg\min_{\mathbf{c}\in\mathbb{C}^{2K+1}}\|\mathbf{B}_s\,\mathbf{c} - \mathbf{f}_s\|_2,
 \label{eq:fourier-recon}
\end{equation}
computed via \texttt{torch.linalg.lstsq}. The reconstructed field at all $N_\varphi$ target locations is then $\hat{\mathbf{f}}=\mathrm{Re}(\mathbf{B}_t\,\mathbf{c})$, where $\mathbf{B}_t\in\mathbb{C}^{N_\varphi\times(2K+1)}$ evaluates the same basis at the full grid. No explicit conjugate-symmetry constraint is imposed; the real part is taken as the final reconstruction, which is sufficient for real-valued plasma fields.

\subsection{Conditioning and contiguous-subset limitations}

When the $N_s$ sparse samples are uniformly distributed around the full period (e.g.\ every $(N_\varphi/N_s)$-th plane), the basis matrix $\mathbf{B}_s$ inherits the orthogonality of the DFT and is well conditioned. In the present work, however, the \gridzsubset\ training planes are contiguous (indices $0,\ldots,\gridzsubset-1$), covering approximately $\gridzsubset/\gridz$ of the toroidal domain. 
